\documentclass[aip,rsi,reprint,amsmath,amssymb,a4paper,floatfix]{revtex4-2}
\def\pdftitle{MagTox}
\def\authorname{Leon Abelmann}
\def\pdfsubject{}
\def\pdfkeywords{}
\def\pdfbackref{none}

\usepackage[pdftex,
        colorlinks=true,
        urlcolor=darkblue,               % \href{...}{...}
        anchorcolor=darkblue,
        filecolor=green,                   % \href*{...}
        linkcolor=darkblue,              % \ref{...} and \pageref{...}
        menucolor=darkblue,
        citecolor=darkblue,
        pagebackref,
        backref={\pdfbackref},
        pdfpagemode={UseOutlines},
        bookmarks=true,
        bookmarksopen=true,
        pdftitle={\pdftitle},
        pdfauthor={\authorname}, 
        pdfsubject={\pdfsubject},
        pdfkeywords={\pdfkeywords}
        ]{hyperref}

\usepackage{graphicx}
\DeclareGraphicsExtensions{.pdf,.png,.jpg}
\usepackage{thumbpdf}
\usepackage{color}
\definecolor{darkgreen}{rgb}{0,0.5,0}
\definecolor{darkblue}{rgb}{0,0,0.5}
\definecolor{brown}{rgb}{0.98,0.92,0.73}
\definecolor{red}{rgb}{1,0,0}
\definecolor{yellow}{rgb}{1,1,0}
\definecolor{blue}{rgb}{0,0,1}
\definecolor{green}{rgb}{0,1,0}
\definecolor{purple}{rgb}{1,0,1}
\definecolor{gray}{rgb}{0.8,0.8,0.8}
\definecolor{black}{rgb}{0,0,0}
\definecolor{white}{rgb}{1,1,1}
\definecolor{gold}{rgb}{1.,0.84,0.}

\usepackage{amsmath}
\usepackage{wasysym}
\usepackage{booktabs}

\usepackage{pdfsync}

\usepackage{siunitx}

\usepackage{nicefrac}

\def\bs{\boldsymbol}

\def\bs{\boldsymbol}

\newcommand{\grad}[1]{%
        \bs{\nabla} #1
}

\usepackage[paper=a4paper,total={170mm,241mm},top=20mm,left=20mm,columnsep=8mm]{geometry}\usepackage{siunitx}
\usepackage{nicefrac}

\def\figurewidth{0.8\columnwidth}
\def\widefigurewidth{\columnwidth}

\newif\ifcmtr
\cmtrtrue
\ifcmtr %  Highlight comment (and hide it when necessary)
\newcommand{\cmtr}[1]{ %
   [\color{red} \textbf{#1} \normalcolor]%
}%
\else
\newcommand{\cmtr}[1]{ %
}%
\fi

\newif\ifcmtrj
\cmtrjtrue%
\ifcmtrj %  Highlight comment (and hide it when necessary)
\newcommand{\cmtrj}[1]{ %
   [\color{green} \textbf{#1} \normalcolor]%
}%
\else
\newcommand{\cmtrj}[1]{ %
}%
\fi

\begin{document}
\title{Permanent magnet systems to study the interaction between
  magnetic nanoparticles and cells in microslide channels}

\date{\today}
\author{Leon Abelmann$^{1,2}$}
\author{Eunheui Gwag$^{1,3}$}
\author{Baeckkyoung Sung$^1$}
\affiliation{ 
  $^1$KIST Europe, Saarbr\"ucken, Germany\\
  $^2$Currently at Delft University of Technology, The Netherlands\\
  $^3$Currently at School of Chemical and Biological Engineering, Institute of Chemical Processes, Seoul National University, Republic of Korea}

\begin{abstract}
  We optimized designs of permanent magnet systems to study the effect
  of magnetic nanoparticles on cell cultures in microslide
  channels. This produced two designs, one of which is based on a
  large cylindrical magnet that applies a uniform force density of
  \qty{6}{MN/m^3} on soft magnetic iron-oxide spherical nanoparticles
  at a field strength of over \qty{300}{mT}. We achieved a force
  uniformity of better than \qty{14}{\percent} over the channel area,
  leading to a concentration variation that was below our measurement
  resolution. The second design was aimed at maximizing the force by
  using a Halbach array. We indeed increased the force by more than
  one order of magnitude at force density values over
  \qty{400}{MN/m^3}, but at the cost of uniformity. However, the
  latter system can be used to trap magnetic nanoparticles efficiently
  and to create concentration gradients. We demonstrated both designs
  by analyzing the effect of magnetic forces on the cell viability of
  human hepatoma Hep\,G2 cells in the presence of bare Fe$_2$O$_3$ and
  cross-linked dextran iron-oxide cluster-type particles
  (MicroMod). Python scripts for magnetic force calculations and
  particle trajectory modeling as well as source files for 3D prints
  have been made available so these designs can be easily adapted and
  optimized for other geometries.
\end{abstract}

\maketitle % Insert title

\tableofcontents

\clearpage

\section{Introduction}

Depending on their concentration, micro- and nanoparticles can be
cyto-toxic~\cite{Elsaesser2012} and can enter our bodies accidentally. They can also
be administered intentionally in biomedical procedures such
as drug delivery~\cite{Anselmo2016,Mornet2004} and \textit{in vivo}
imaging.\cite{Pankhurst2003, Zabow2011}

To assess the effect of nanoparticles on cells, most interaction
studies start with tests on \textit{in vitro} cell cultures in
multi-well plates. In the case of small molecules, diffusion ensures
that the molecule concentration is reasonably constant over the volume
of the well. However, micro- and nanoparticles are subject to
sedimentation. This has two implications. First, sedimentation
gradually increases particle concentration at the cell membrane, the
rate of which depends strongly on the particle diameter. Secondly, the
particles exert a force on the cell membrane, which may affect
particle incorporation.\cite{Teeguarden2006}

The material composition of the nanoparticles is the prime aspect
determining particle toxicity. In this study we focus on magnetic
nanopartices, in particulare those composed of iron-oxide. Iron-oxide
has adverse effects at a concentration in the order of
\qty{50}{\micro g/mL}, far above the \qty{1}{\micro g/mL} of for
instance silver nanoparticles.\cite{VazquezMunoz2017, Greulich2012}
The advantage of magnetic nanoparticles however is that they can be
manipulated by external magnetic fields. The magnetic forces that one
can apply are several orders of magnitude greater than gravitational
forces. Therefore, by magnetically attracting nanoparticles towards
the bottom of the well, we can accelerate sedimentation and increase
particle incorporation. This study of the relation between force and
toxicity can help us to better understand the toxicity of non-magnetic
nanoparticles.

The ability to manipulate magnetic nanoparticles magnetically has made them useful 
for targeted drug delivery,\cite{Pondman2014, Kim2010}
mechano-stimulation~\cite{Kilinc2016} and hyperthermia
treatment.\cite{Colombo2012} Unprotected iron-oxide particles have
strong effects on cell viability. Experiments with epithelial cells
showed that viability noticeably declined above \qty{50}{\micro g/mL}.\cite{Rafieepour2019} Similarly, fibroblast cell proliferation
strongly decreased in the presence of bare iron-oxide nanoparticles with
concentrations above \qty{50}{\micro g/mL}.\cite{Berry2003} Therefore,
magnetic nanoparticles for biomedical treatment are usually embedded
in a protective coating.\cite{Soenen2010} Their interaction with cells
depends strongly on the type of coating. For instance, starch-coated particles  show no reduction of cell viability up to concentrations of
\qty{500}{mg/mL}. In contrast,
dextran-sulfide-coated particles have noticeably decreased cell viability
at concentrations above \qty{50}{\micro g/mL}.\cite{Yanai2012}

The application of magnetic forces has a strong effect on cell viability.
To study the effect of magnetic forces, permanent magnets are typically
positioned below the cell culture dishes or multi-well plates. Prijic
and colleagues elegantly demonstrated the increased uptake of
super-paramagnetic particles in human melanoma and mesothelial cell
cultures placed on top of magnets.\cite{Prijic2010} They found
that the total iron content in the cell, measured by inductively
coupled plasma atomic emission spectroscopy, increased by a factor of
\num{3}--\num{8}. Cell viability decreased by \qty{50}{\percent} for
a concentration of \qty{100}{\micro g/ mL}.

A particularly convincing method to study particle uptake is to use
magnetic particles to transinfect cells, a method called
magnetofection.\cite{Haim2004, Scherer2002} Pickard and
Chari~\cite{Pickard2010} attached green fluorescent protein (GFP)
plasmids to Neuromag SPIONs. When the particles entered the cell, the
plasmids were reproduced. The subsequent generation of GFP determined
whether cells are transfected. The application of force by means of
magnetic field gradients enhanced the uptake by a factor of 5. A slow
oscillation appeared to have a positive effect.

Particles in a cell can be identified with optical microscopy by means
of fluorescence, for which Dejardin and colleagues used
ScreenMag-Amine magnetic particles tagged with
fluorescein.\cite{Dejardin2011} The particles were treated with
activated penetratin to increase their uptake. By integrating the
intensity of the emitted light over the sample area, an increase in
uptake of about \qty{30}{\percent} was observed. A similar approach was
taken by Venugopal and colleagues~\cite{Venugopal2016} as well as by
Park and colleagues,\cite{Park2016} who used flow
cytometry to demonstrate an increase in uptake ranging from a factor of 
\num{0.5} to \num{7}, respectively.

In all cases cited above, the magnets were smaller than or of a similar size as the
cell culture area, leading to particle accumulation in the center of
the observation area and subsequent loss of information on particle
concentration. In this work, we designed magnetic systems for maximum
uniformity so that the particle concentration is known more precisely.

If a uniform particle concentration is not important, for instance if
the aim is merely to capture particles, then
non-uniform forces are not an issue. In that case, higher gradients
can be achieved by arrays of magnets. For example, a (circular)
Halbach array has been used to capture magnetically labeled connective
tissue progenitor cells from bone marrow~\cite{Joshi2015} or to trap
magnetically labeled cells in the leg~\cite{Riegler2011} or the
brain.\cite{Shen2017} Using a yoke with an embedded rotating magnet,
so that the field can be removed~\cite{Hoffmann2002}, magnetically
labelled polyclonal antibodies could be captured at efficiencies of \qty{95}{\percent}.~\cite{Gomes2018}

In a microfluidic system, non-uniform magnetic forces are used to
capture magnetically labeled circulating tumor cells by
means of small single magnets~\cite{Kang2012} or by
linear~\cite{Stevens2021} or rotating cylindrical Halbach
arrays.\cite{Xue2019}

Most of the magnets used are alloys of NdFeB. These can be manufactured
down to millimeter dimensions. Smaller magnets can be obtained with thin-film technology by using lithography to pattern the thin film. Examples
are 12-$\mu$m-thick, \qtyproduct{40x40}{\micro m} permalloy squares, which
were used to capture cells decorated with magnetic
particles,\cite{Ooi2013} or 100-nm-thick \qtyproduct{300x50}{\um} bars
 to grow magnetically labeled
neurons directly.\cite{Alon2015}

The research mentioned above focused primarily on the effect of magnetic
forces on cells, but less so on the design of the magnetic
systems themselves. In this contribution, we therefore analyzed the magnetic force profiles in
more detail, either to achieve as
uniform a field as possible by using bigger magnets, or to achieve high
forces by using Halbach arrays. The designs were made specifically for
microslides dedicated to cell cultures. We used these combinations of
magnetic systems and microslides to study the effect of magnetic
forces on the viability of human liver cells in the presence of
magnetic nanoparticles.

In the following, we introduce models to calculate magnetic
fields, forces and trajectories of magnetic nanoparticles. We compare
calculations and measurements of the fields and forces on a system
designed for maximum uniformity and on a system based on a Halbach
array. We will then draw conclusions regarding the strength of fields and forces of
both magnetic systems and the efficiency of particle capture under
flow conditions for a Halbach array.

To illustrate the application to cell viability studies, we analyzed
the effect of bare and protected iron-oxide particles on the viability
of liver cells with and without an applied magnetic force.

%%% Local Variables:
%%% mode: latex
%%% TeX-master: "../paper"
%%% End:

\section{Theory}

\subsection{Field and force calculations}
\label{sec:FieldForceCalculations}
% Cades for cylindrical
% %Calculations using MagMMEMS
To calculate the magnetic field and forces generated by a single
cylindrical magnet, we integrated magnetic charge densities. In
contrast to finite-element methods, the field is calculated only at
the points of interest, which is much faster at high precision. The
resulting equations are generated automatically by the MagMMEMS
package, which is a preprocessor for Cades.\cite{Delinchant2007}  The
input files are available in the Supplementary Material.
%Figure~\ref{fig:Field}, Figure~\ref{fig:FzFxAreaInterest}

For Halbach arrays, the calculated field profiles serve as input to
particle trajectory calculations. Therefore, we implemented the
integrals in Python using the \texttt{dblquad} integration method of
the \texttt{scipy} package. The source files are available on
github (\href{https://github.com/LeonAbelmann/Trajectory}{https://github.com/LeonAbelmann/Trajectory}).

% Python for halbach, omdat we trajectories willen bepalen.
%Figure~\ref{fig:trajectory}
%Figure~\ref{fig:capture}

The force on a magnetic object, which is small compared to the spatial
variation of the externally applied field $\bs{B}$~[T], can be
approximated from its total magnetic moment $\bs{m}$~[A/m]
\begin{eqnarray}
  \label{eq:grad}
  \bs{F} = -\grad{(\bs{m}\bs{B})}
  \text{.}
\end{eqnarray}
%
% Force
The magnetic moment of a magnetic object in a fluid is generally a
function of strength, direction and history of the applied field. The
applied field is a combination of the external field and the field of
all other magnetic particles in the fluid. Moreover, very small
particles will be subject to Brownian motion. Therefore, in principle, the
calculation of forces on magnetic particles in a magnetic field
gradient is complex. To obtain first approximations, we consider a
single particle that is either a permanent magnetic dipole or a soft
magnetic sphere.

%\subsection{Permanent magnet approximation}
In the case of the permanent magnetic dipole approximation, we assume a particle
with a permanent magnet moment $\mu_0m_\text{r}=I_\text{r}V_\text{p}$,
where $I_\text{r}$ [T] is the remanent magnetization of the particle
with volume $V_\text{p}$ [m$^3$]. We further assume that field changes are
slow such that particles can rotate against viscous drag into the
direction of the field. In this case, Eq.~(\ref{eq:grad}) reduces
to
\begin{eqnarray}
  \label{eq:magnet}
  \bs{F}=m_\text{r} \grad{B}\text{, } B=|\bs{B}| \text{.}
\end{eqnarray}

%\subsection{Linear permeability approximation}
For the second approximation, we assume a soft
magnetic sphere with a susceptibility of $\chi$, which is the ratio between
the magnetization $I$~[T] in the particle and the internal field
$B_\text{in}$~[T]. As a sphere has a demagnetization factor of
1/3,  the internal magnetic field is 
\begin{equation*}
  \label{eq:chi}
  B_\text{in}=-B + \frac{1}{3}I = -B + \frac{1}{3}\chi
  B_\text{in} 
  \text{,}
\end{equation*}
where
\begin{equation*}  
  I = \frac{3\chi}{(3+\chi)}B 
\end{equation*}
and all fields are (anti-)parallel. In this approximation, the energy
and resulting force are
\begin{equation}
\begin{aligned}
  \label{eq:3}
  U &= -\frac{1}{2}V_\text{p}\frac{3\chi}{\mu_0(3+\chi)}B^2  \\
  F &= \frac{1}{2}V_\text{p}\frac{3\chi}{\mu_0(3+\chi)} \grad{B^2}  \\
  &=   V_\text{p}\frac{3\chi}{\mu_0(3+\chi)} (\bs{B}\grad{})\bs{B} 
  \text{.}
\end{aligned}
\end{equation}
The factor 1/2 originates from integrating from $-\infty$, where the
energy is 0, and we used the vector identity
$ \grad{B^2}=2 (\bs{B}\grad{})\bs{B}$.

\subsection{Trajectory calculations}
\label{sec:trajectories}
% Refereer naar paper Michiel Stevens

In our experiments with the Halbach array, we filled the channel with the
nanoparticle suspension when it is on top of the array. The
nanoparticles are taken along by the fluid flow and at the same time
dragged down towards the array. As the particles are small and
flow velocities are low, inertial effects are negligible. In that case,
the magnetic force $F_\text{mag}$ (N) is instantly balanced by the
drag force
\begin{equation}
F_\text{mag} = 6 \pi \eta r (v_\text{p} - v_\text{f})
\text{,}
\end{equation}
where $\eta$ is the fluid viscosity (set to \SI{1}{mPa s}). From the
force balance, we can obtain the velocity of the particles
$v_\text{p}$ with respect to the fluid background velocity $v_\text{f}$
(m/s).

% Channel is filled, so there will be flow. Assume a parabolic flow profile
The average background fluid velocity $v_\text{flow}$ is estimated
from the filling time and length of the channel to be around
\SI{25}{mm/s}. As the flow velocity is zero at the surfaces of the
channel, the flow profile is parabolic
\begin{equation}
v_\text{f}(z) = 1.5v_\text{flow} \left(1-(\frac{2z}{h}-1)^2 \right)
\text{,}
\end{equation}
where $h$ is the channel height (\SI{0.8}{mm}). We assume the flow direction to
be uniquely along the channel length ($x$). However, the magnetic force is
allowed to have components in all directions. 

The resulting differential equation is solved by using the
\texttt{solve\_int} routine of the \texttt{scipy.integrate} package
using the RK23 integration method with an absolute tolerance of
\num{1e-5} and relative tolerance of \num{2e-3}. \cite{Midelet2017,Stevens2021} Particles at the top of the
channel have the least chance of being captured. In case of incomplete
capture, the capture height and capture efficiency are obtained by
integrating backwards in time from a trajectory that starts exactly at
the bottom of the channel at the channel exit. Source codes are 
available on github
\href{https://github.com/LeonAbelmann/Trajectory}{https://github.com/LeonAbelmann/Trajectory}.

%%% Local Variables:
%%% mode: latex
%%% TeX-master: "../paper"
%%% End:

\section{Experimental}

\subsection{Fabrication cylindrical magnet system}
\begin{figure}
    \centering
    \includegraphics[width=\widefigurewidth]{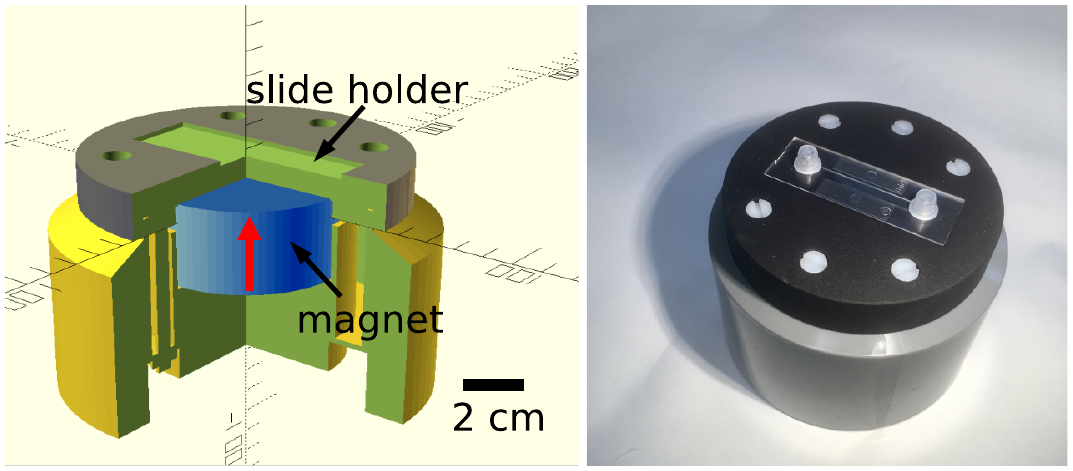}
    \caption{System to investigate the effect of a high magnetic field
      gradient from a single magnet on the interaction of magnetic
      nanoparticles on cells. The magnet has a diameter of \qty{70}{mm}
      and a height of \qty{35}{mm}. The cell culture is located \qty{10.5}{mm} above the magnet in a slide channel.}
    \label{fig:IbidiHolder}
\end{figure}

The aim of the first magnetic system was to obtain as uniform a force as
possible over the area of interest. To maximize uniformity, we
used the largest NdFeB magnet we could readily obtain
(Supermagnete.de). This magnet has a diameter of \qty{70}{mm}, a height
of \qty{35}{mm} and is made of N45, which is specified to have a
remanent magnetization of \qty{1.3}{T}. These magnets can seriously
injure the experimentalist if accidentally brought too close
together. Therefore we encased them in PVC (\qty{70}{mm} (121605),
\qty{100}{mm} (121457) and \qty{125}{mm} (121620) reducer rings from
Wildkamp, Netherlands). The position of the $\mu$-slide channel and
the distance to the magnet were accurately fixed using a 3D-printed
nylon plastic top holder, see Figure~\ref{fig:IbidiHolder}. The
holder can be disinfected by a \qty{70}{\percent} isopropanol
solution. The source files for the 3D-printed holder are available in
the Supplementary Material.

\subsection{Fabrication Halbach magnet system}
\begin{figure}
    \centering
    \includegraphics[width=\figurewidth]{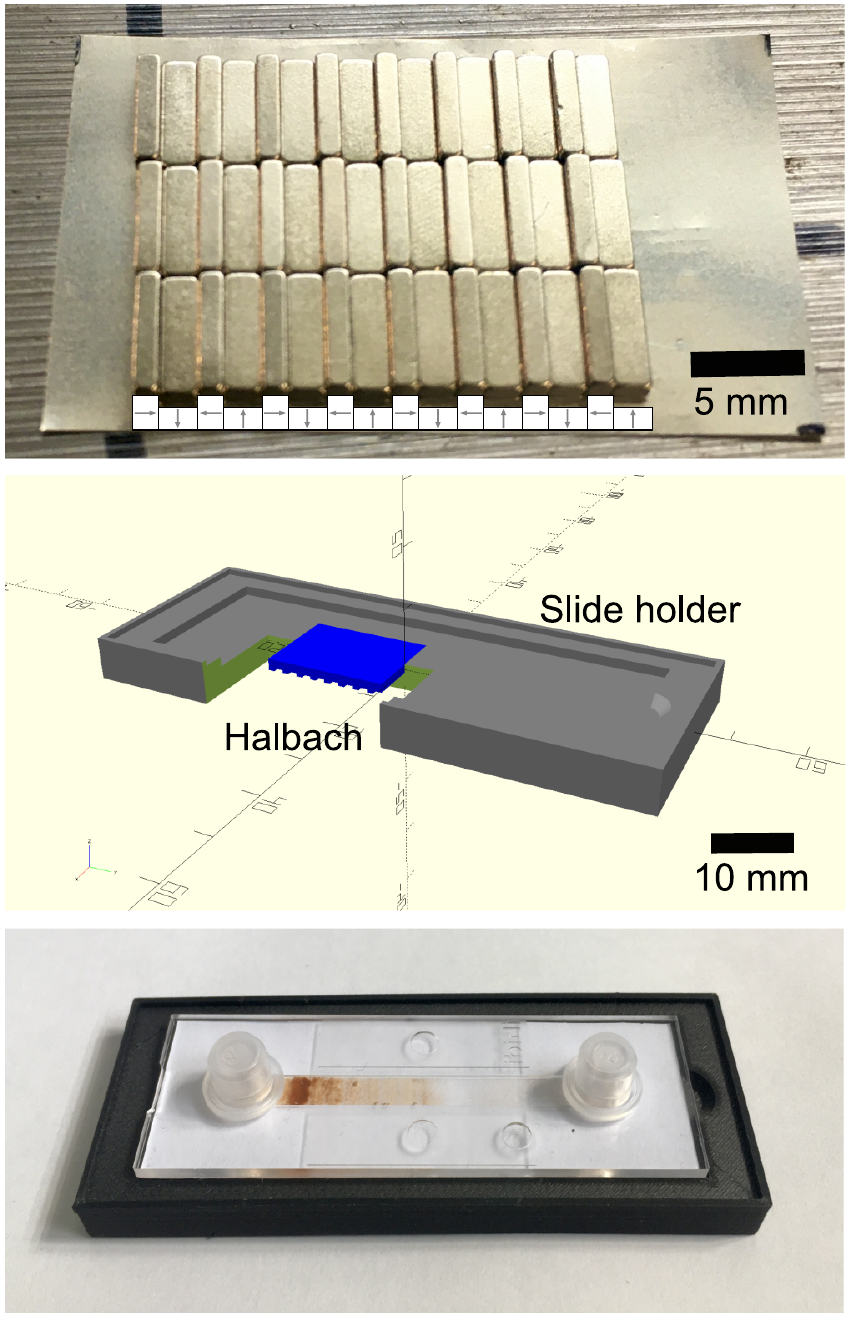}
    \caption{System to investigate the effect of concentration
      gradients. A strong field gradient is achieved by means of a Halbach array consisting of \num{48}
      \qtyproduct{1.5x1.0x5.0}{mm} magnets (top). The Halbach array (top) is
      embedded in a 3D-printed holder (center) that keeps the channel
      slide in place (bottom). }
    \label{fig:HalbachIbidi}
\end{figure}

The Halbach array was assembled from \num{48} individual
\qtyproduct{1.5x1.0x5.0}{mm} magnets (supermagnet.de), see
Figure~\ref{fig:HalbachIbidi}. The magnets were placed at a distance on a thin
copper foil on a tapered soft magnetic plate 
and then carefully pushed together. After assembly, the array was glued
together using a thin cyanoacrylate glue, and carefully slid off the
thin end of the wedge. The array, including the copper foil, was then
glued with the foil on top onto a 3D-printed holder. The design for
the holder is available in the Supplementary Material.

\subsection{Magnetic field measurements}
The magnetic field above the cylindrical magnet
(Figure~\ref{fig:Field}) and Halbach array
(Figure~\ref{fig:FieldGradient}) was measured with a MetroLab THM1176
three-axis Hall sensor. The sensor was placed at the same height as
the channel of the $\mu$-slide and translated using a manual
micromanipulator. The location of the probe was estimated to within
\qty{1}{mm} and was then fine-tuned laterally to the symmetry in
the field profile. The orientation of the probe with respect to the
vertical ($z$-axis) is difficult to adjust and could deviate up to \ang{4}.

The field profile of the Halbach array was visualized using a magnet
viewer, see Figure~\ref{fig:FieldGradient}, which has low contrast when
the field is parallel to the sheet (TRU components 507706).

\subsection{Microscopy}
Overview images of the particle distribution in the $\mu$-slide
channel were taken by a Canon EOS 800D camera with a Canon EFS18-55
lens at a distance of approximately \qty{20}{cm}, see Figure~\ref{fig:Halbachgradient}.

The contrast variation over the channel is very low as observed on the cylindrical magnet. For this reason, we mounted the camera in a
light-shielded box and illuminated it with a home-built LED ring driven
by a constant voltage, see Figure~\ref{fig:Gradient5Hours}. For future
experiments, we advise using a constant-current driver to avoid
illumination intensity changes during long experiments. Time-lapse
images were taken at 2-min intervals using the external shutter
input of the Canon camera. The images were then isolated using an optocoupler and driven
by an Arduino Uno.

Images were taken using the raw CR2 format (\qty{41}{MB} each) until
the 64-GB SD card was full (approximately \num{128} images). A Python
script was used to load the CR2 images and average the pixel intensity $I$
for the green channel over \num{160} slices of \qtyproduct{16x400}{pixel}
 (\qtyproduct{0.2x3.6}{mm}). A reference image was taken from an
empty channel ($I_0$), and the measured average intensity was converted to
absorbance using $\log_{10}(I/I_0)$.

% Stereo microscope
Higher-resolution images of particle suspension inside the $\mu$-slide
channel on top of the cylindrical magnet were taken with a Zeiss
Stemi 508 stereo microscope, see Figure~\ref{fig:ParticlesUniform}. 

% Inverted microscope,
When the $\mu$-slides are removed from their magnet, they can be
imaged by transmitted light microscopy. The images in
Figure~\ref{fig:ParticlesGradient} were taken with a Zeiss Axiovert
100 using a \qty{12}{V} \qty{100}{W} halogen light source, a Zeiss LD
A-Plan 10$\times$/0.25 Ph1 lens and a Jenoptik Gryphax Prokyon
camera. The microscope stage is equipped with stepper motors. To image
along the channel, images were taken every \qty{1}{mm} at fixed
illumination and combined into a \qtyproduct{0.7x36}{mm} image using
a Hugin photo stitcher (hugin.sourceforge.io). The intensity
profiles were extracted from the stitched images by means of Gwyddion's
``extract profile along arbitrary lines'' tool, using the green
channel for the bare Fe$_2$O$_3$ particles and the red channel for the
core-shell particles and a width of 128~pixels.

% Cell imaging
The $\mu$-slide channels are dedicated to cell studies. Images of
cells (Figures~\ref{fig:MicroscopyUniform}, \ref{fig:MortalityGradient}
and ~\ref{fig:MicroscopyGradient}) with a size of
\qtyproduct{2048x2048}{pixel} were taken with a Leica DMi8
fluorescence microscope with a K5 sCMOS camera. Both a \num{20}$\times$ and a
\num{40}$\times$ lens were used, calibrated at \num{324} and
\qty{162}{nm/pixel}, respectively. To observe the channels on the Halbach array, we 3D-printed a holder that allowed us to image 
the location
along the channel with a reproducibility of at least \SI{1}{mm}.

\subsection{Particles}
We used both bare Fe$_2$O$_3$ and core-shell magnetic nanoparticle
suspensions. The bare Fe$_2$O$_3$ suspension was based on iron(III)
oxide powder (Alfa Aesar NanoArc, 45007). According to the
manufacturer, the particles have a diameter of \qtyrange{20}{40}{nm}
and are over \qty{98}{\percent} in the $\gamma$ crystal phase.  The
powder was mixed with demineralized water at an iron concentration of
\qty{10.6}{\mg /\ml}.  From experiments by Prijic~\cite{Prijic2010}
and Rafieepour~\cite{Rafieepour2019}, we esimated that for significant
cell mortality, we needed to apply concentrations well above
\qty{250}{\ug /ml}. . Therefore, for cell studies on the cylindrical
magnet, the stock suspension was diluted \num{27} times into a
phosphate-buffered saline (PBS) solution with a pH of \num{7.4} to
achieve an iron concentration of \qty{393(20)}{\ug /\ml},
corresponding to (\num{2} to \num{18})$\times$\qty{e12}{particles /
  \ml}. For the experiments on the Halbach array, the suspension was
diluted further to \qty{100(5)}{\ug /\ml}, so that the expected
maximum concentration was approximately equal to the experiment with
the cylindrical magnet.

In addition to the bare Fe$_2$O$_3$ powder, we used crosslinked
dextran iron-oxide cluster-type particles prepared by a core-shell
method (Bionized NanoFerrite (BNF) 94-00-102 from Micromod). These
particles are red fluorescent (redF) and shipped in a PBS
suspension. According to the manufacturer, they have a core of
\qtyrange{75}{80}{\percent} (w/w) magnetite and a shell of
dextran. The particles have a reported diameter of \qty{100}{nm} with
a magnetite crystallite diameter of about \qty{20}{nm}. The
hydrodynamic diameter of these particles lies between \num{95} and
\qty{140}{nm}.\cite{Giustini2009} The original iron concentration as
reported by the manufacturer is \qty{6.0}{\mg /\ml}, and particle
concentration is \qty{6e12}{particles/mL}. The original suspension was
diluted into PBS to achieve an iron concentration of \qty{280(10)}{\ug
  /\ml}, corresponding to a concentration of \qty{3e12}{particles /
  \ml}.

Suspensions were kept inside a refrigerator in the dark. Before use, the
tubes with suspensions were vigorously shaken and ultrasonically agitated
for approximately one minute. For studies that did not
involve cells, the PBS was replaced with demineralized water.

\subsection{Channel slides}
All experiments were performed using Ibidi $\mu$-Slide I Luer channels
with an Ibitreat surface coating to promote cell adhesion (Ibidi
80606). The dimension of the channels is \qtyproduct{0.8x5.0x50}{mm}.

\subsection{Cell staining}
Cells were stained for the microscopy images in
Figures~\ref{fig:MicroscopyUniform} and~\ref{fig:MicroscopyGradient}
with an Iron Stain Kit (ab150674; Abcam, Cambridge,
UK). To identify cellular uptake and localization of iron-oxide
nanoparticles, HepG2 cells in the $\mu$-slide were washed with PBS and
incubated in a mix of potassium ferrocyanide and hydrochloric acid for
3~min. After being rinsed with deionized water, the cells were
counterstained with a nuclear fast red solution.

\subsection{Mortality assay}
For the cell studies (Figures~\ref{fig:MortalityUniform} and
\ref{fig:MortalityGradient}), human hepatoma HepG2 cells (ATCC,
HB-8065) were cultured in Eagle's minimum essential medium
supplemented with \qty{10}{\percent} fetal bovine serum and
\qty{1}{\percent} penicillin-streptomycin in an incubator at
\qty{37}{\celsius} and \qty{5}{\percent} CO$_2$ atmosphere. The cell
concentration was determined by means of a hemocytometer and diluted
to \qty{5e4}{cells/\ml}. Of this solution, \qty{200}{\ul} was
introduced into the $\mu$-slide channels. The maximum cell
density was \qty{40}{cells/mm^2}.

The Ibidi channels were left inside the incubator for \qty{24}{hours}
on top of the holder with and without a magnet before analysis. To
assess cell viability, we used a live/dead double-staining assay
(Sigma-Aldrich, 04511).  Ten fluorescent
microscopy images (Leica DMi8) were chosen randomly over the channel
area for analysis. Average and standard deviations were calculated from three
independent experiments (\num{30} images in total).

%%% Local Variables:
%%% mode: latex
%%% TeX-master: "../paper"
%%% End:

\section{Results and discussion}

\subsection{Magnetic field and forces}
To verify the calculations of fields and forces, we measured the
fields of both the cylindrical magnet and the Halbach array as well as
the concentration variation over the channel filled with magnetic
nanoparticles.

\subsubsection{Uniform forces: Cylindrical magnet}
The force field above the permanent cylindrical magnet varies with
distance to the magnet surface in both strength and direction. For
experiments with magnetic nanoparticles inside the channel slide, we
want the lateral forces in the plane of the channel to be as small as
possible, yet the vertical force to be as high as possible and very
uniform over the area of interest.  Using the soft sphere model of
Eq.~\ref{eq:3}, we determined that the optimal height of the
channel slide for both conditions is \qty{10.5}{mm} above the magnet.

\begin{figure}
    \centering
    \includegraphics[width=\figurewidth]{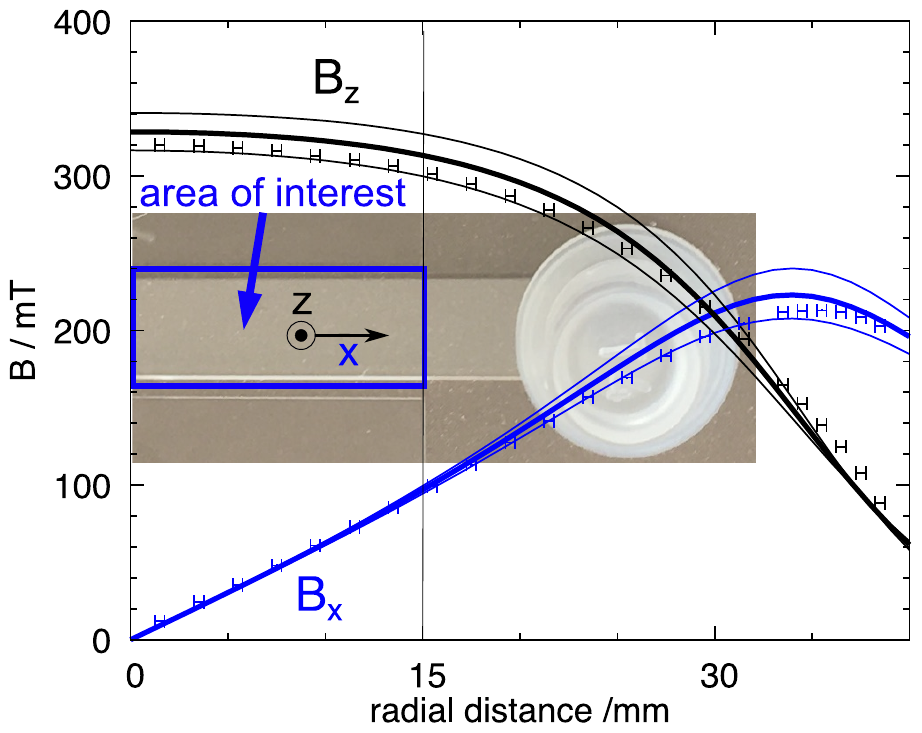}
    \caption{Field strength above the cylindrical magnet showing both
      calculated and measured vertical (top) and lateral (bottom)
      field components at the height  of the channel slide
      (\qty{10.5}{mm}). Calculated field values for \qty{9.5}{mm} and
      \qty{11.5}{mm} are shown by thinner lines as well. Assuming a remanent
      magnetization of \qty{1.30}{T}, the field is predicted within
      measurement error. Over the length of the channel (\qty{30}{mm}),
      the vector component of the field in the vertical direction is
      dominant, with a strength greater than \qty{300}{mT}.}
    \label{fig:Field}
\end{figure}

Figure~\ref{fig:Field} shows the measured and calculated magnetic
field components perpendicular to channel $B_\text{z}$ and along
channel $B_\text{x}$ at the optimum height of
\qty{10.5}{mm}. The measurements follow the prediction quite
accurately, but the deviations are greater than the field measurement
uncertainty. Additional sources of uncertainty are the angle of the
sensor with respect to the magnet's surface and the distance between
the sensor and the magnet's surface. The latter deviation has a greater
impact and may vary over the radial distance. Therefore, we also show
predictions that are \qty{1}{mm} higher and lower than
the optimum height, respectively. Assuming a remanent magnetization of \qty{1.3}{T}
as provided by the manufacturer's data, the measurements fall in this
band within measurement error. Over the length of the channel
(\qty{30}{mm}, radial distance \qty{15}{mm}), the perpendicular field
component is dominant with a value in excess of \qty{300}{mT}. The
field strength varies only by \qty{5}{\percent}, and the field angle
rotates at a maximum of $\pm$\qty{17}{\degree}.

\begin{figure}
    \centering
    \includegraphics[width=\figurewidth]{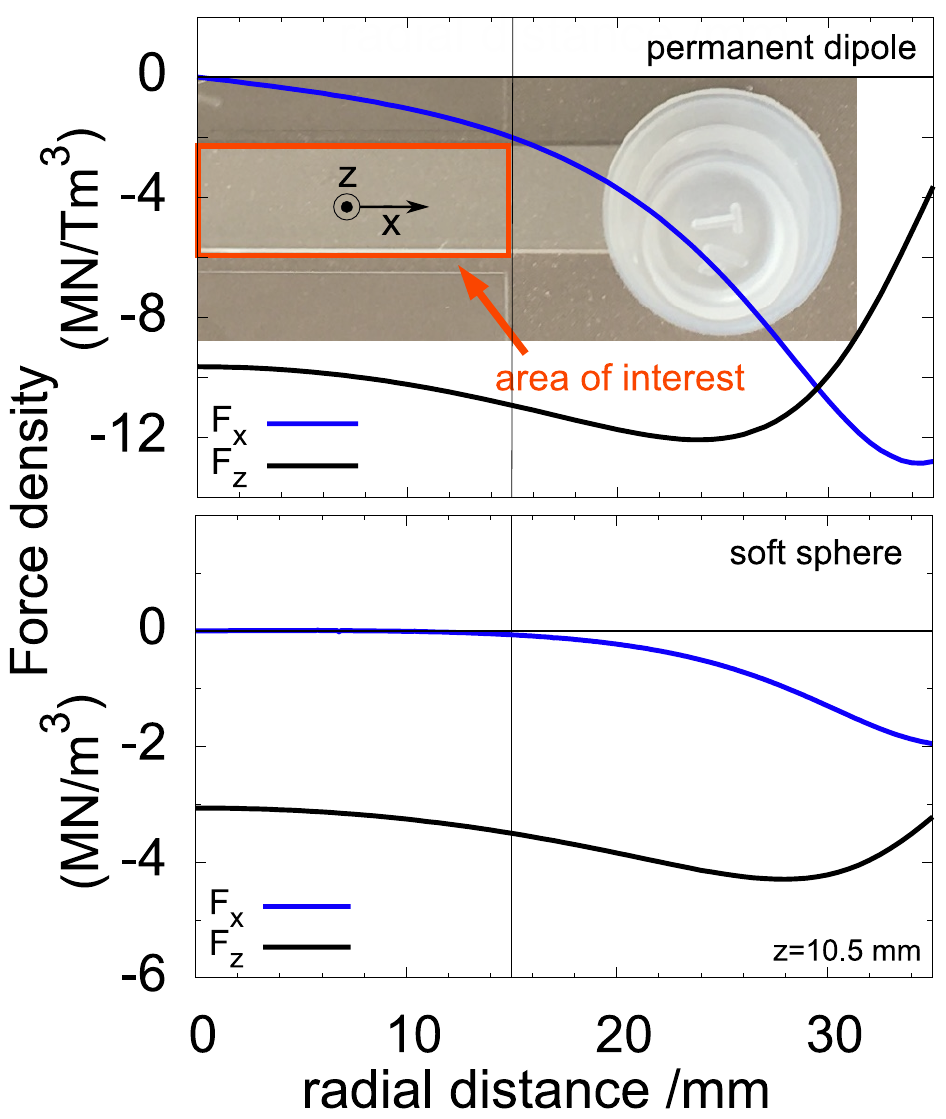}
    \caption{Force densities calculated from the field profiles shown in
      Figure~\ref{fig:Field} in vertical (black) and lateral (blue)
      directions as a function of the position over the length of the
      channel slide. We apply two extreme models. The top curve shows
      the result for a model in which we assume that the particle is a
      permanent magnet with remanent moment $I_\text{r}$
      (Eq.~\ref{eq:magnet}). To obtain the force, multiply by
      $V_\text{p}I_\text{r}$. For the bottom curve, we assumed that the
      particle is a sphere of volume $V_\text{p}$ and susceptibility
      $\chi$ (Eq.~\ref{eq:3}). To obtain the force, multiply
      the value on the vertical axis by $3V_\text{p}\chi/(3+\chi)$.
      The vector component of the force in the vertical direction is
      dominant, with a strength in excess of \qty{3}{MN/m^3}. Over the
      length of the \qty{30}{mm} channel, the force strength remains
      within \qty{14}{\percent} and the force direction varies by less
      than \qty{1}{\degree}.}
    \label{fig:FzFxAreaInterest}
\end{figure}

Figure~\ref{fig:FzFxAreaInterest} shows the calculated forces on particles for
both the permanent dipole magnet (top) and soft sphere (bottom)
models.  For the permanent dipole model, the forces are normalized to
the particle magnetic moment $I_\text{r} V_\text{p}$ [Tm$^3$]. A
typical particle has a magnetization in the range from \qty{0.1}{}
to \qty{1}{T}, so force densities are on the order of
\qty{1}{} to \qty{10}{MN/m^3}. For comparison, the gravitation force density on a
typical iron-oxide particle is  only \qty{40}{kN/m^3} (standard gravity
times mass density difference with water). 

For the soft sphere model, the forces are normalized to
$3V_\text{p}\chi/(3+\chi)$, which ranges from \qty{0}{} to
\qty{3}{}$V_\text{p}$. A typical iron-oxide particle has a susceptibility
 greater than 1~\cite{Yun2014a}, so force densities are
in the same range as for the permanent dipole approximation.

Both models show a variation of \qty{14}{\percent} in the vertical
component of the force over the length of the 30-mm channel. The
soft sphere model has a much weaker lateral force. At the entrance of
the channel, the force tilts only about \qty{1}{\degree} inward,
whereas for the permanent dipole model, it tilts at \qty{10}{\degree}.

\begin{figure}
    \centering
    \includegraphics[width=\figurewidth]{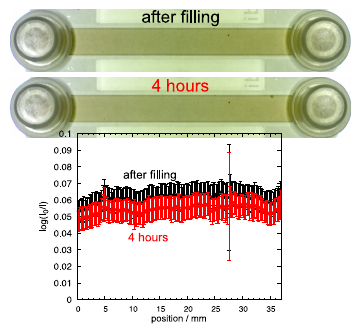}
    \caption{Channel filled with core-shell magnetic nanoparticles immediately after filling (top, black) and after \qty{4.3}{h}. The
      graph shows the absorbance averaged over a 3.6-mm band in
      the center of the channel, referenced to an empty channel. In
      the Beer--Lambert approximation, the absorbance is proportional to
      the concentration. Error bars indicate the standard deviation of
      the absorbance over the width of the band. The concentration
      variation over the length of the channel is less than the
      measurement uncertainty of \qty{12}{\percent}. The reduction in
      light absorbance with time could also be caused by a gradual
      decrease in the intensity of the light source. A time-lapse video of the
      256-min process  is available in the Supplementary
      Material.}
    \label{fig:Gradient5Hours}
\end{figure}

\begin{figure}
    \centering
    \includegraphics[width=\widefigurewidth]{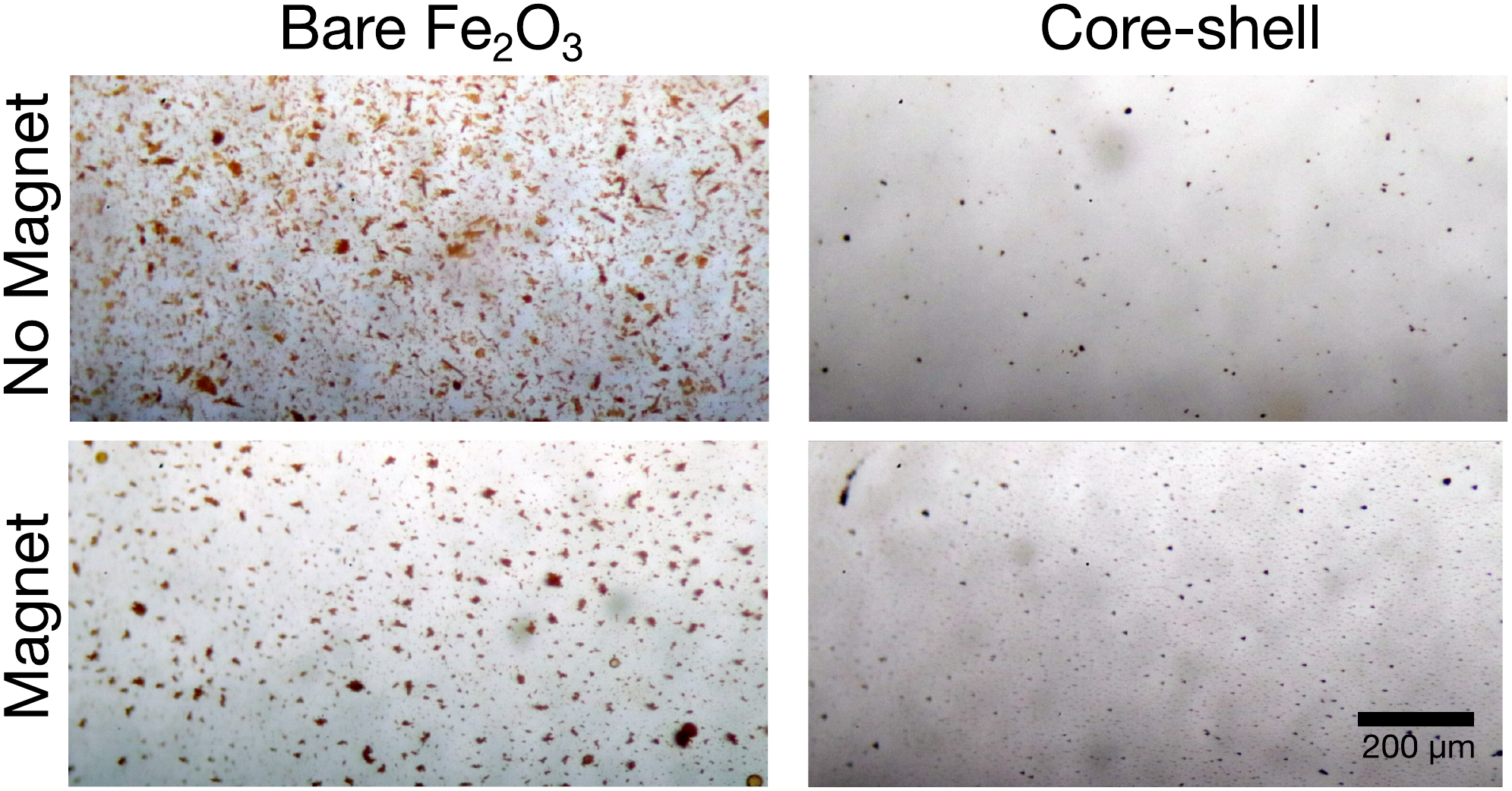}
    \caption{Optical microscopy of bare Fe$_2$O$_3$ and core-shell
      particles without an applied magnetic field (top) and on top of
      the cylindrical magnet. In the presence of the magnetic field,
      the particles have a greater tendency to cluster, leading to a reduction of
      light brown areas in the case of Fe$_2$O$_3$ and the appearance
      of many more small spots in the case of the core-shell
      particles.}
    \label{fig:ParticlesUniform}
\end{figure}

The large magnet produces a very uniform distribution of magnetic
particles. Figure~\ref{fig:Gradient5Hours} shows a channel slide
filled with a diluted suspension of 5-nm iron-oxide
particles. No concentration gradient can be observed even after more
than \qty{4}{hours}. We measured the intensity of the image over a
3.6-mm band over the center of the channel. By comparison with
an empty channel, we can measure the concentration
variation. Assuming that the logarithm of the ratio of the filled and
unfilled channels is proportional to the concentration (Beer--Lambert
law), the variation in concentration along the channel is less than
the measurement uncertainty of \qty{12}{\percent}. A time-lapse video of the
256-min process with timesteps of \qty{32}{s} is
available in the Supplementary Material
(\texttt{CoreShell.mp4}). There is a gradual reduction in overall
intensity. Closer inspection of the images shows that the magnetic
field causes the appearance of many small spots in the case of
core-shell particles, see Figure~\ref{fig:ParticlesUniform}, left. We
assume that the appearance of these spots is caused by particle
clustering, which may explain the reduction in overall reflection
of light. Clustering can also be observed for the bare Fe$_2$O$_3$
particles, see Figure~\ref{fig:ParticlesUniform}, right.

\subsubsection{Concentration gradients: Halbach array}
%Figures~\ref{fig:FieldGradient}, \ref{fig:Halbachgradient},
%\ref{fig:trajectory},\ref{fig:capture}, \ref{fig:ParticlesGradient}

The aim of the large cylindrical magnet described above was to create as
uniform a force as possible. If uniformity is not an issue, then a
Halbach array can create much higher  forces. 

\begin{figure}
    \centering
    \includegraphics[width=\figurewidth]{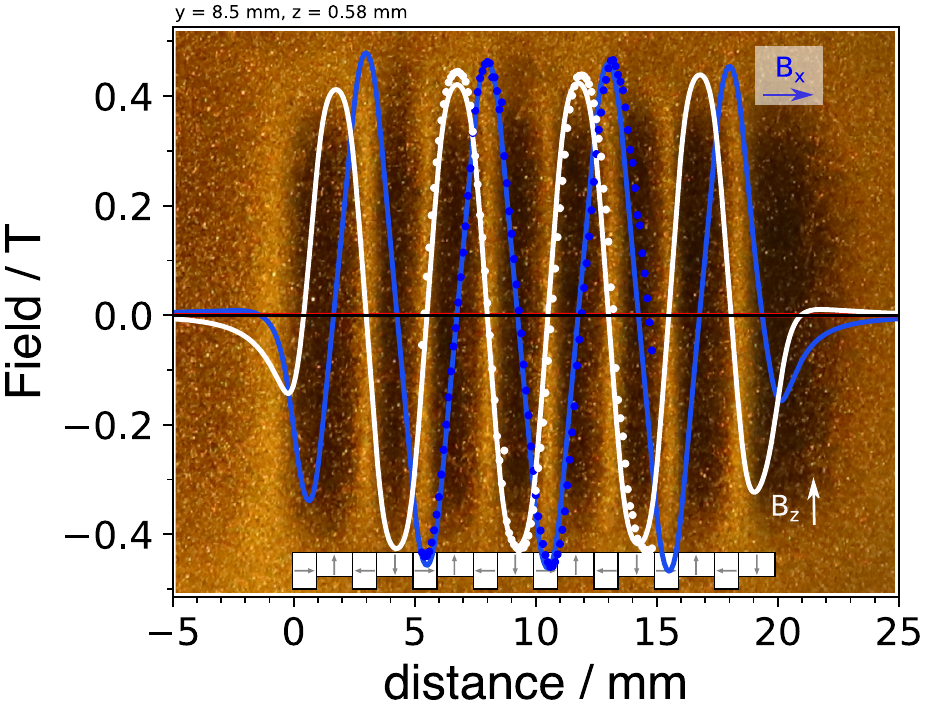}
        \includegraphics[width=\figurewidth]{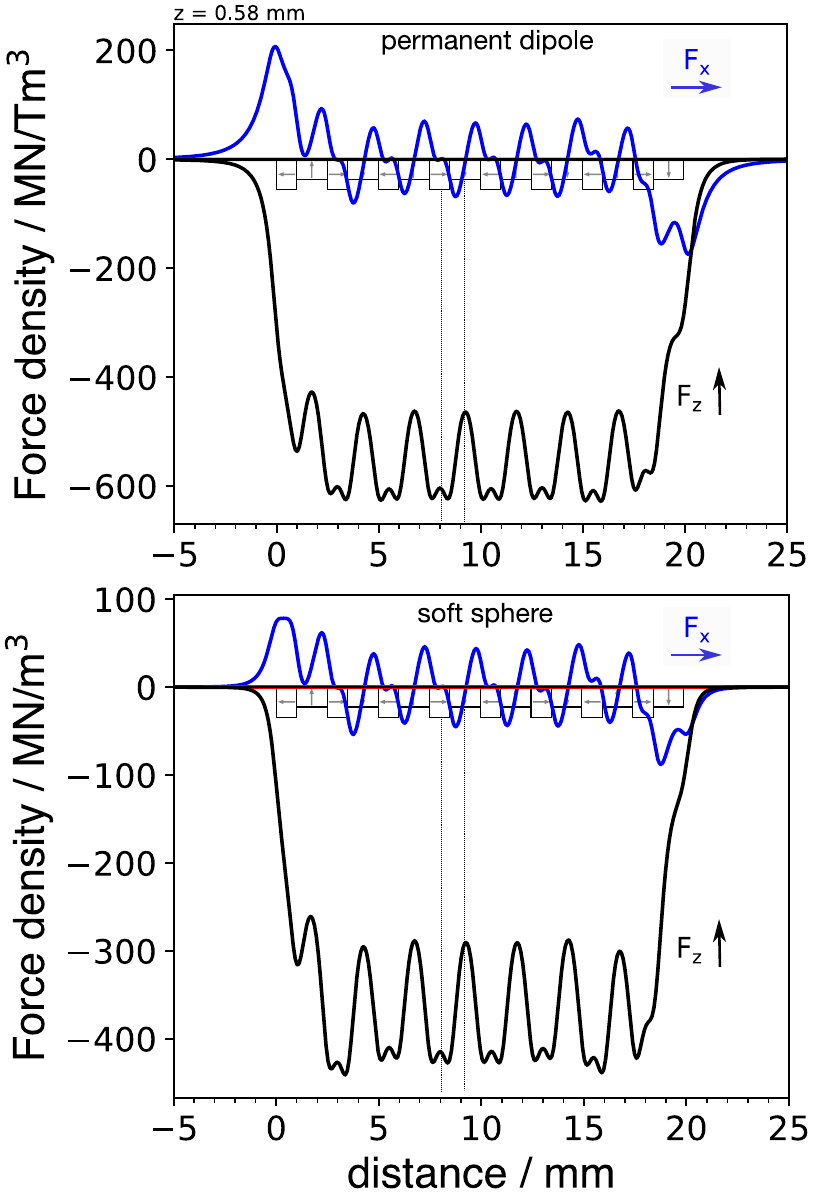}
        \caption{Calculated and measured vertical and lateral
          components of the field (top) and calculated forces on
           permanent magnet (center) and soft magnetic sphere (bottom)
          particles as a function of distance along the Halbach
          array. To obtain the force, multiply by
          $V_\text{p}I_\text{r}$ for the permanent magnet or by
          $3V_\text{p}\chi/(3+\chi)$ for the soft magnetic sphere.  In
          the background of the top field graph, the field direction
          is shown by means of a magnetic viewer (in the light areas, the
          field is parallel to the viewer sheet). The circles are
          measured values. The height was set to \qty{0.58}{mm} in the
          calculation to match these values. As can be seen by the
          shift in the measured pattern, the dimensions of the magnets
          are slightly larger than specified. The field strength is of
          the same order as was the case for the uniform field in
          Figure~\ref{fig:Field}. However, the forces are  more than one
          order of magnitude higher because the individual magnets in
          the Halbach array are \num{70} times smaller. In contrast to
          the cylindrical magnet, the Halbach array has very
          non-uniform forces, with lateral forces being both positive and
          negative and vertical forces varying by as much as
          \qty{30}{\percent}.}
    \label{fig:FieldGradient}
\end{figure}

Figure~\ref{fig:FieldGradient} (top) shows the calculated and measured
magnetic field profiles of a Halbach array of \num{16}
\qtyproduct{1.5x1.0}{mm} magnets. The total width of the array (in
$y$-direction) is \qty{15}{mm}. The values of the field components fit
very well with measurements if we assume a distance between probe and
magnet surface of \qty{0.58}{mm}, which is very reasonable. The
maximum field strength is on the order of \qty{400}{mT}, which is
comparable to the large cylindrical magnet.

An image of a magnetic viewer film is shown in the background. The
light and dark areas indicate regions where the field is parallel and perpendicular to the plane
of the viewer film, respectively. The measurement and model show very good
qualitative as well as quantitative agreement. There is a small shift in the
pattern, which may be caused by the fact that the distance between
magnet centers is slightly increased because of the glue used to
assemble the array.

Using this model, we can calculate the forces on magnetic
nanoparticles. The bottom two graphs in Figure~\ref{fig:FieldGradient}
show the in-plane and perpendicular force components for a model,
assuming that the nanoparticles are either permanent magnets (center) or soft
magnetic spheres (bottom). The perpendicular force is always
attractive, and varies by \qty{25}{\percent} (permanent dipole) to
\qty{30}{\percent} (soft sphere). In contrast, the in-plane force can
be both positive and negative. Particles will therefore tend to
diffuse to regions where the in-plane force is zero, leading to very
non-uniform particle concentrations. However, the force densities are
approximately \num{60} to \num{90} times higher than for the
cylindrical magnet. This is the effect of  reducing the size of the
magnets by a factor of \num{70} and using the Halbach configuration.

\begin{figure}
  \centering
  \includegraphics[width=\widefigurewidth]{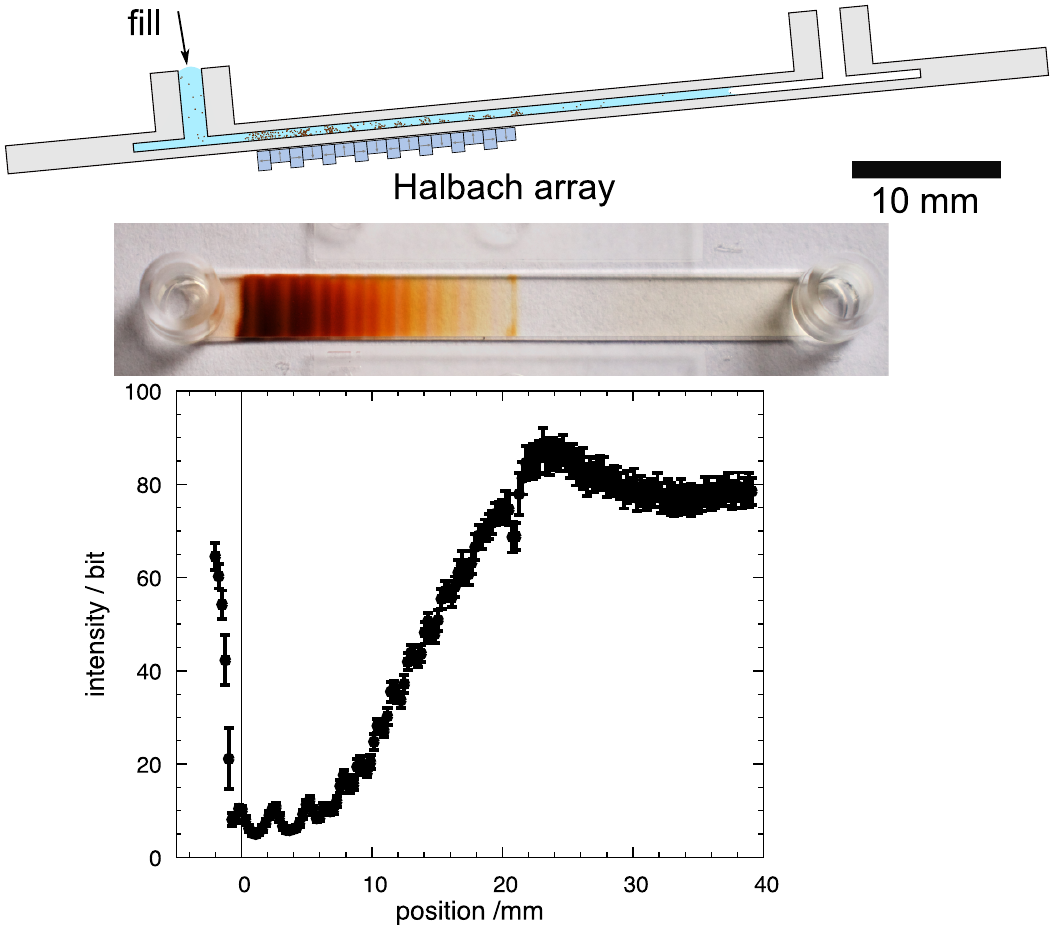}
  \caption{As the channel slowly fills, bare Fe$_2$O$_3$
    nanoparticles are captured in the stray field of the Halbach
    array.  Particles are captured as the fluid moves over the
    array, leading to a concentration gradient. The contrast of the
    image is shown below and increases almost linearly from
    \qty{5} to \qty{90}{bit} in the range of
    \qtyrange{8}{20}{mm}. On the basis of the Beer--Lambert law, we
    conclude that there is
    a logarithmic reduction in
    concentration in this region. A video of the filling process
    (\texttt{GradientFilling.mov}) is available in the Supplementary
    Material. }
    \label{fig:Halbachgradient}
\end{figure}

In the experiment with the large cylindrical magnet, the channel was
filled first with magnetic nanoparticles, so that the concentration
distribution was uniform. The channel slide was then carefully lowered vertically
along the axis of the magnet to avoid disturbing the concentration. In
the case of the Halbach array, we first positioned the empty slide on
the array, slightly titled, and filled it from the lower side, see
Figure~\ref{fig:Halbachgradient}.  As the suspension flows over the array, particles are trapped by the magnetic field
gradient, and the concentration decreases. This achieves a gradient in the
concentration. The filling process can be observed
in the video in the Supplementary Material
(\texttt{GradientFilling.mov}).

As the particles are trapped at the transitions between the
magnets, the concentration oscillates slightly. However, the increase in
intensity is more or less linear over a range of
\qty{12}{mm} (from \qtyrange{8}{20}{mm}). Assuming a Beer--Lambert law for the
relation between intensity and concentration, the concentration
decreases approximately exponentially. This suggests that a
fixed fraction of the particles is indeed captured as the suspension flows
over the array.

\begin{figure}
    \centering
    \includegraphics[width=\figurewidth]{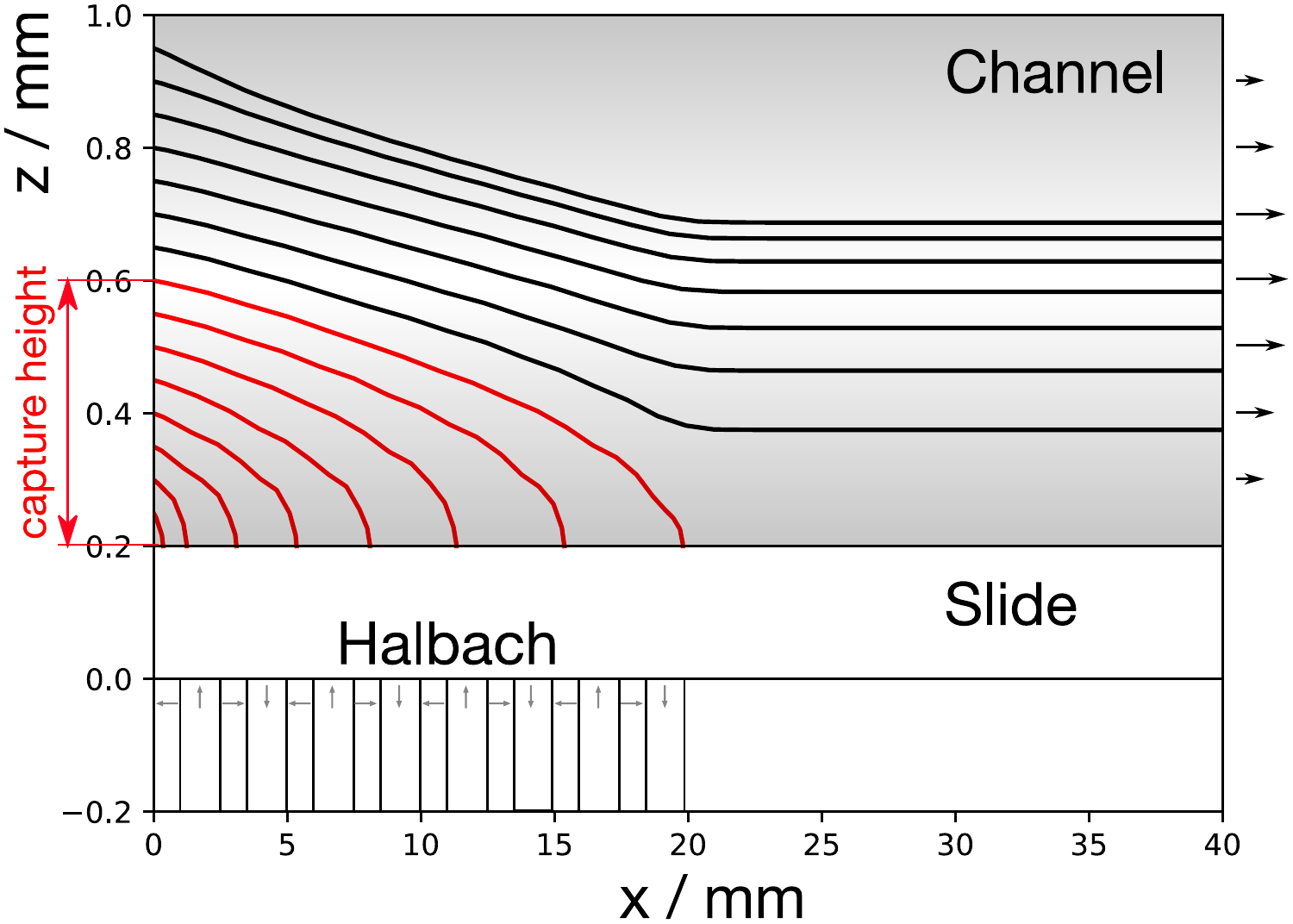}
    \caption{Calculated trajectories of magnetic nanoparticles
      released at the left-hand side of the Halbach array. The flow profile
      in the channel is assumed to be of parabolic shape with an average
      flow velocity of \qty{25}{mm/s}, or \qty{6}{mL/min}. Note that
      the $x$ and $z$ scales of the image are very different. The
      particular example we have chosen here illustrates the situation
      where \qty{50}{\percent} are captured (particles with a radius
      of \qty{67}{nm} and magnetic remanence of \qty{1.0}{T}.)}
    \label{fig:trajectory}
\end{figure}

The velocity of the particles with respect to the fluid is determined
by the balance between magnetic and drag force. Using the magnetic
force calculated by the permanent dipole model and assuming a
parabolic flow profile, we can calculate the trajectories of magnetic
nanoparticles in suspension, see
Section~\ref{sec:trajectories}. Figure~\ref{fig:trajectory} shows an
example of magnetic nanoparticles with a radius of \qty{67}{nm} and a
remanent magnetization of \qty{1.0}{T} in a flow with an average
velocity of \qty{25}{mm/s}. For this particular situation, about half
the particles are trapped on the array, whereas the other half leaves
the channel slide. Particles at the bottom of the channel are captured
more easily. Using this type of calculation, we can predict a capture
height and capture efficiency. In this particular calculation, the
capture height is \qty{0.4}{mm} and the capture efficiency is
\qty{50}{\percent}.

\begin{figure}
    \centering
    \includegraphics[width=\figurewidth]{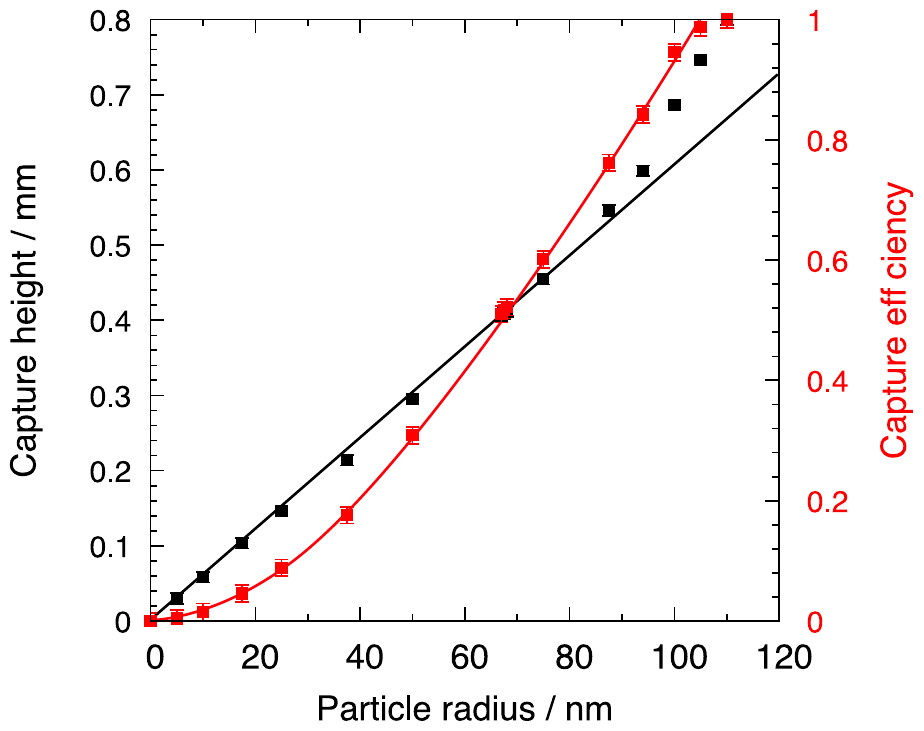}
    \caption{Calculated capture height, see
      Figure~\ref{fig:trajectory} (left axis, black) and resulting capture efficiency
      (right axis, red) as a function of particle
      radius. The capture rate for particles with
      radii below \qty{20}{nm} is less than \qty{7}{\percent}.  Owing to
      the parabolic flow profile, the capture efficiency increases
      quadratically with increasing particle radius. A capture
      efficiency of \qty{50}{\percent} is reached for a particle radius of
      about \qty{67}{nm}, which is the example shown in
      Figure~\ref{fig:trajectory}. For small particle radii, the
      capture height increases linearly with increasing particle
      radius. Above this value, the linear relationship between
      capture height and radius is lost. The red line is not a fit,
      but shown merely to guide the eye. Full capture is achieved for
      particle radii above \qty{110}{nm}.}
    \label{fig:capture}
\end{figure}

The magnetic force of the particles in the suspension scales with
their volume, whereas the drag force scales with the radius. The smaller
the particles, the less chance they have of being
captured. Figure~\ref{fig:capture} shows the estimated capture height
and capture efficiency as a function of particle radius (again
assuming particles with a remanent magnetization of \SI{1.0}{T}). Up
to a particle radius of about \qty{75}{nm}, the capture rate increases
linearly with particle radius. Owing to the parabolic flow profile,
the capture efficiency increases quadratically. Particles with radii
above \qty{110}{nm} are all captured. Particles with radii of less
than \qty{20}{nm} have a chance of less than \qty{7}{\percent} of
being captured in this specific situation.

\begin{figure}
    \includegraphics[width=0.5\textwidth]{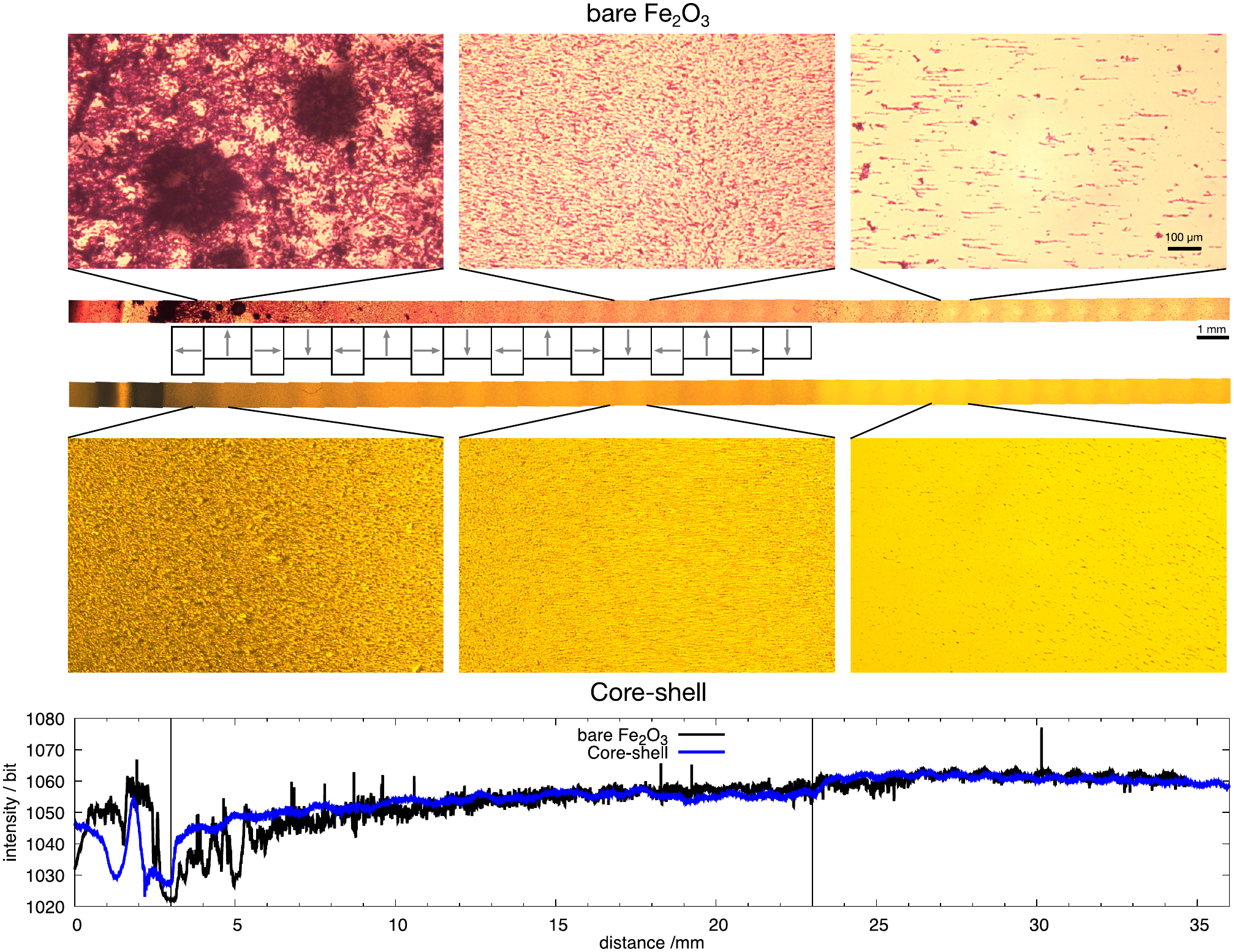}
    \caption{Microscopy images of bare Fe$_2$O$_3$ (top) and
      core-shell particles (bottom) after  the channel slide is removed
      from the Halbach array. The images were stitched to visualize a
      \num{36}$\times$\qty{0.7}{mm^2} band along the center of the
      channel and are available in full resolution in the
      Supplementary Material (\texttt{FullGradient*.png}). Typical
      examples are extracted, and the estimated location of the
      Halbach array is shown. The bare Fe$_2$O$_3$ particles show a more
      non-uniform distribution. Particles tend to cluster into
      structures that can be as long as \qty{100}{\um}. At the start of
      the Halbach array, the clusters are irregularly
      distributed. Midway through the array, the clusters align along the
      channel. Not all particles are captured. Those that escape the
      Halbach array tend to align into chains along the channel
      length.  The bottom graph shows intensity profiles along the
      stitched images for both types of particles. The change in
      intensity along the channel is stronger for the bare Fe$_2$O$_3$
      particles. The strong variation at the start of the array is
      caused by the gas bubbles.}
    \label{fig:ParticlesGradient}
\end{figure}

Experiments on liver cell viability were performed with bare
Fe$_2$O$_3$ particles, as described above, as well as with core-shell magnetic
nanoparticles. Figure~\ref{fig:ParticlesGradient} shows stitched
microscope images over the length of the channel slide for both types
of particles. The gradient in particle concentration is clearly
visible. The black dots at the entry of the channel with the
Fe$_2$O$_3$ suspension are gas bubbles. These bubbles did not form in
the suspension of core-shell particles. The Fe$_2$O$_3$ particles show
a less uniform distribution. Rather, the particles cluster into elongated
structures with lengths up to \qty{100}{\um}. At the start of the
Halbach array, these clusters are irregularly distributed. Further
into the channel, the distribution becomes more regular. Above the
in-plane magnets of the array, the clusters align along the channel;
above the perpendicular magnets, the clusters are oriented more
randomly. Not all particles are captured by the Halbach array. At the
end of the array, a number of larger clusters can be found. In the
remaining part of the channel, chains of particles can be seen aligned
very well along the channel length. For the Fe$_2$O$_3$ particles, this
alignment is lost towards the end of the channel. In the case of the
core-shell particles, the alignment turns towards \qty{45}{\degree} at
the end of the channel.

In the case of bare Fe$_2$O$_3$ particles, we observed the formation of
tiny bubbles at the entry of the channel. We have two hypotheses regarding
the origin of these bubbles. They could be caused by heating of the
fluid due to the intense illumination in the microscope. The increased
temperature of the fluid could trigger the release of dissolved gases
(nitrogen, oxygen). Indeed, we have been able to melt the channel
slide by using full-intensity light, so an increase in temperature is
very likely. However, the channel filled with core-shell particles did
not show bubble formation, not even in the very dark, i.e.\ light-absorbing, area from \qtyrange{0}{3}{mm}. Bubbles formed  also at moderate light
intensity, see Supplementary Materials,
video \texttt{Fe2O3.mp4}. Our other hypothesis is that
Fe$_2$O$_3$ causes photo-oxidation of water, releasing hydrogen and
oxygen.\cite{Kay2006} Indeed, when a suspension of bare Fe$_2$O$_3$
is placed in sunlight, gas bubbles are produced continuously for
days. In the case of core-shell particles, the dextran shell may
prevent contact between water and the Fe$_2$O$_3$, which could explain
the absence of bubbles. Further research will be required to
determine the origin of the bubbles. In any case, if the
concentrations are low and/or illumination is moderate, bubbles do
not form.

%%% Local Variables:
%%% mode: latex
%%% TeX-master: "../paper"
%%% End:

\subsection{Cell viability}
We demonstrated the use of two designs on human hepatoma HepG2 cells in
channel slides. The viability of the cells in the presence of magnetic
nanoparticles was observed over a period of one to six days.

\subsubsection{Cylindrical Magnet}
\begin{figure}
    \centering
    \includegraphics[width=\figurewidth]{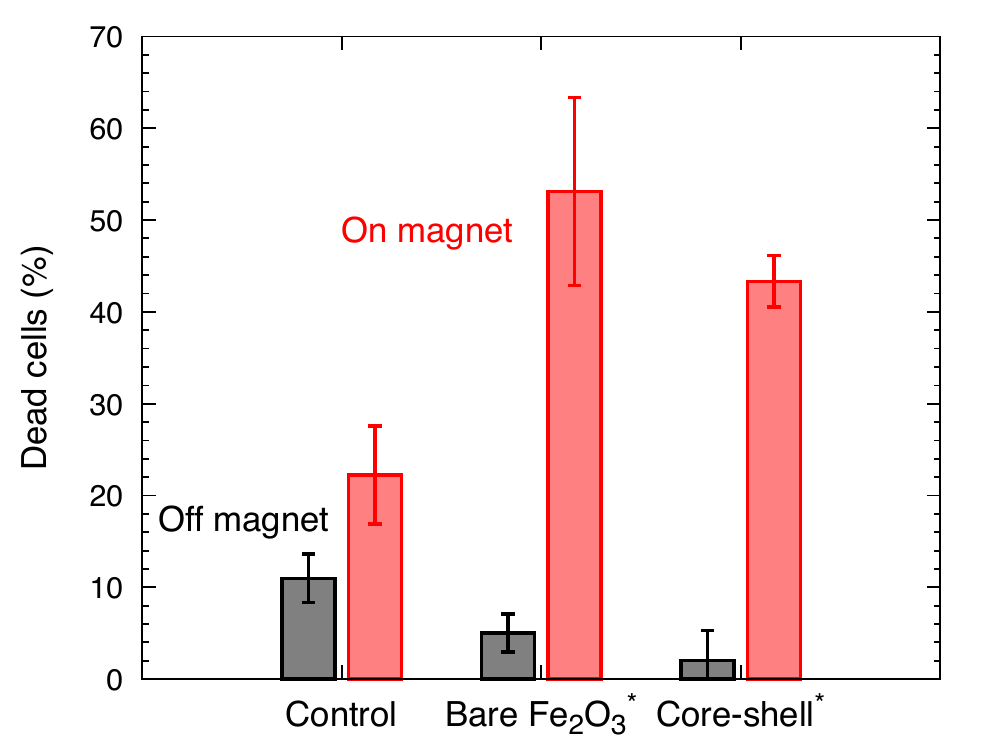}
    \caption{Human hepatoma HepG2 cell mortality after \SI{6}{days},
      with (red) and without (black) application of a uniform magnetic
      field. Error bars indicate the standard error, based on ten
      images per slide and three independent experiments. The control
      without magnetic particles shows that the magnetic field by
      itself has no statistically significant effect on cell mortality
      (confidence less than \SI{95}{\percent}). The presence of bare
      Fe$_2$O$_3$ particles at a concentration of \SI{400}{\micro
        g/mL} and core-shell particles at \SI{280}{\micro g/mL}
      increases cell mortality up to approximately
      \SI{50}{\percent} (confidence exceeds \SI{99}{\percent}).}
    \label{fig:MortalityUniform}
\end{figure}

Figure~\ref{fig:MortalityUniform} shows the viability of HepG2 cells
inside channel slides after six days, with and without application of
a uniform magnetic force and with and without the presence of magnetic
nanoparticles. The application of a magnetic force by itself does not
significantly affect cell viability. However, when there are magnetic
nanoparticles in the solution, the application of a magnetic force
strongly reduces cell viability. The effect is similar for bare
Fe$_2$O$_3$ and core-shell nanoparticles. There seems to be a small
reduction in cell death when magnetic nanoparticles are
added. However, this effect is only statistically significant for the
core shell particles.

\begin{figure}
  \centering
  \includegraphics[width=\widefigurewidth]{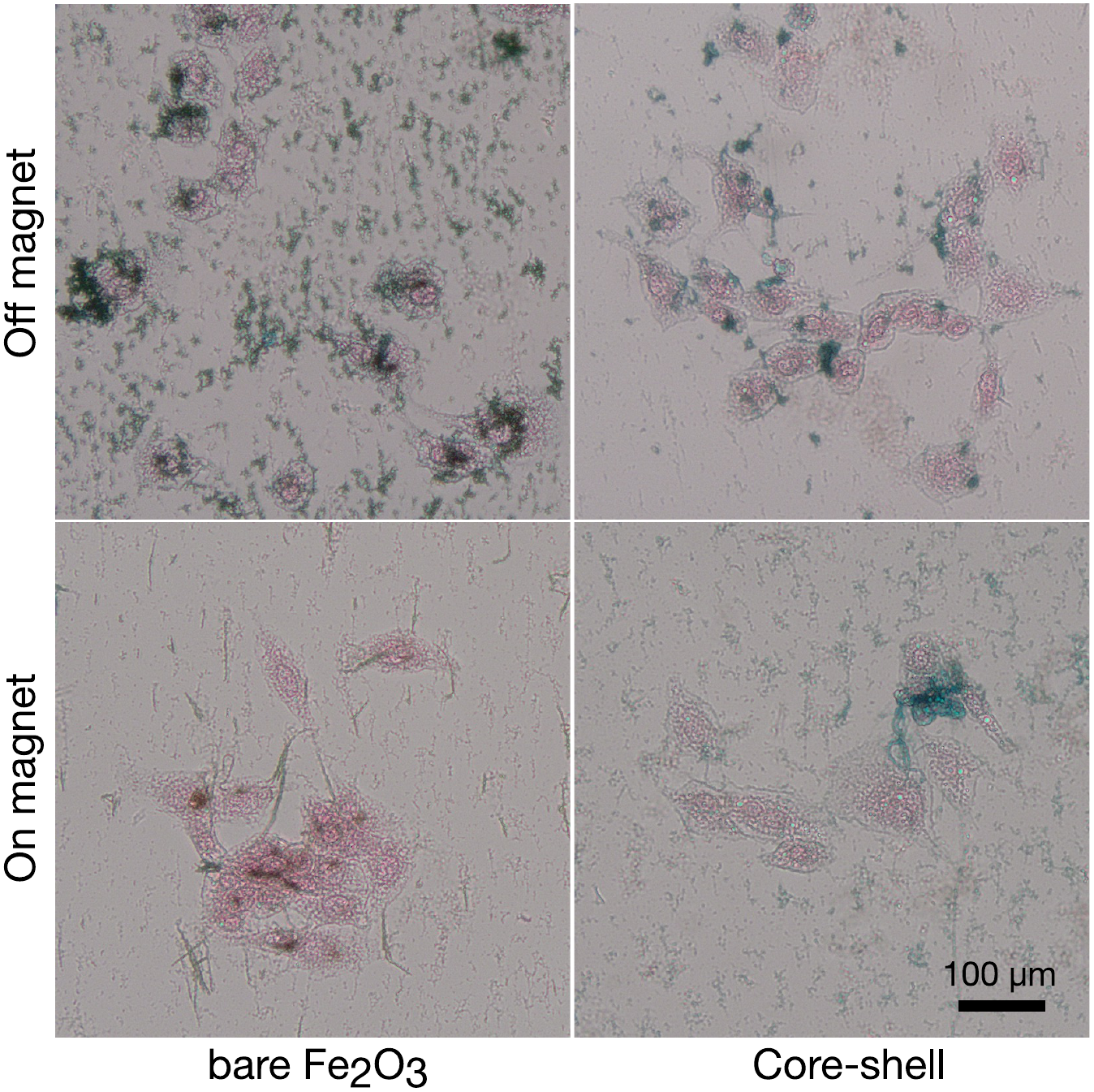}
  \caption{Optical microscopy image of HepG2 cells cultivated for six days in
    the presence of bare Fe$_2$O$_3$ (left, \SI{400}{\micro g/mL}) and
    core-shell particles (right, \SI{280}{\micro g/mL}). The top and bottom rows
    show cells that were cultivated off- and on-magnet, respectively. The nuclei in the cells are stained red; the iron-oxide is stained blue. Cell morphology is not noticeably dependent
    on the type of particles. In the magnetic field, the particles
    agglomerate into chains. This effect is strongest for the bare
    Fe$_2$O$_3$ nanoparticles.}
  \label{fig:MicroscopyUniform}
\end{figure}

Closer inspection of the cells and nanoparticles in the channel slides
shows that, on the magnet, the nanoparticles tend to cluster more strongly, see
Figure~\ref{fig:MicroscopyUniform}. The effect is strongest for the
bare Fe$_2$O$_3$ particles, which form elongated structures with
lengths of up to \qty{100}{\um}. However, cell morphology does not seem
to be strongly affected by the application of a magnetic force. 

%%% Local Variables:
%%% mode: latex
%%% TeX-master: "../paper"
%%% End:

\subsubsection{Halbach array}
\begin{figure}
    \centering
    \includegraphics[width=\figurewidth]{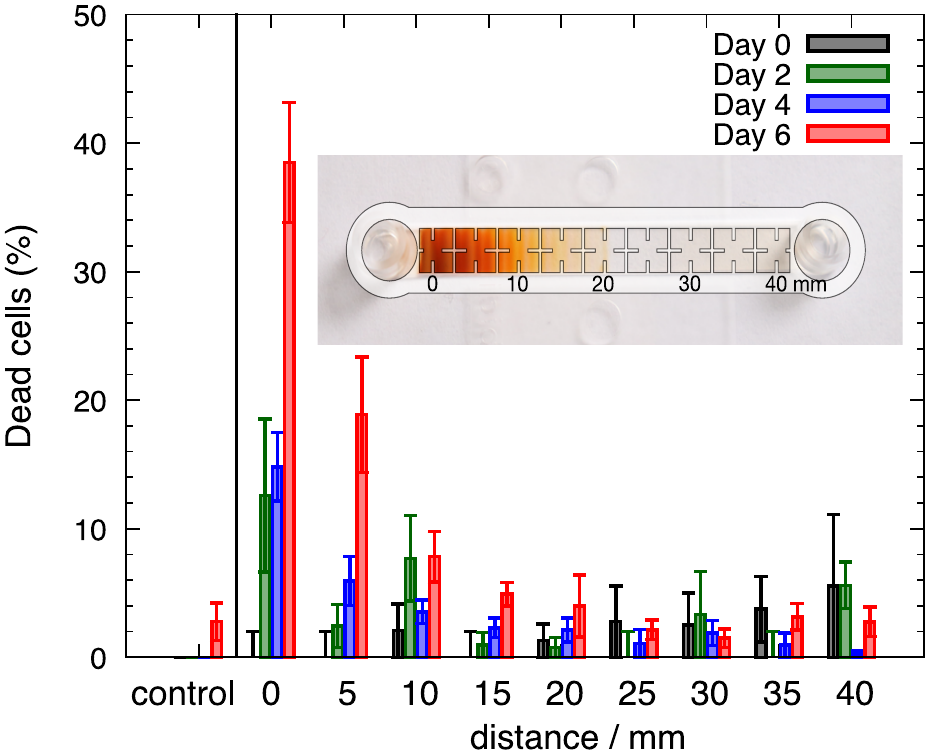}
    \includegraphics[width=\figurewidth]{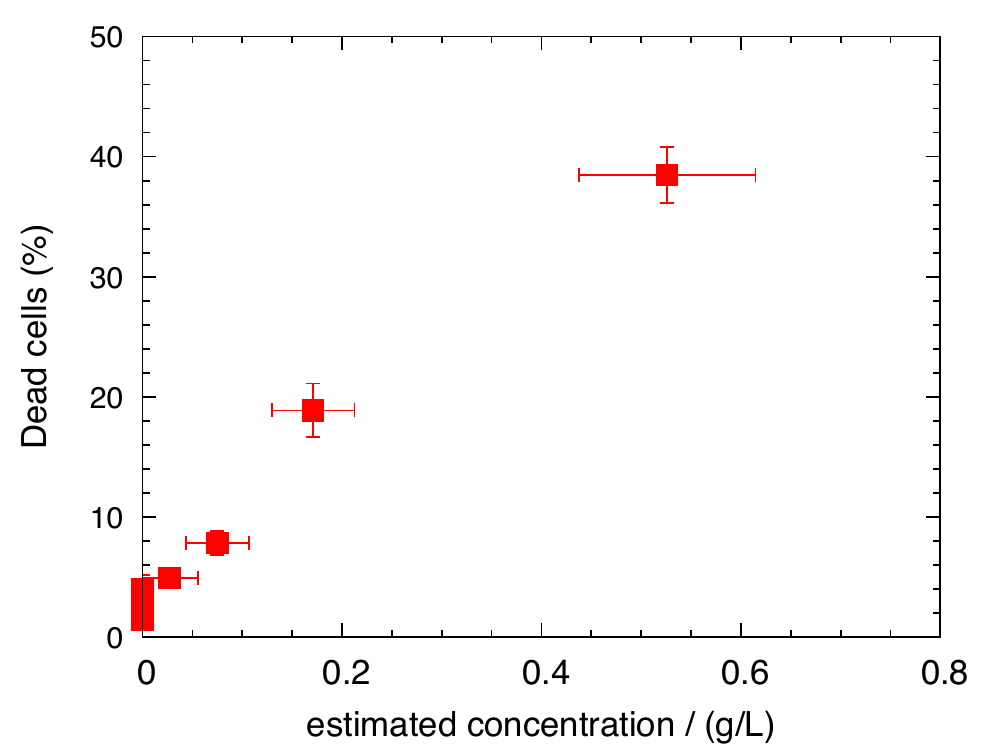}
    \caption{Top: Viability of HepG2 cells cultivated in the presence
      of bare Fe$_2$O$_3$ nanoparticles with concentration decreasing
      to zero at about \SI{20}{mm} as a function of time.  Error bars
      indicate the standard error, based on ten images per slide and
      three independent experiments. As the control (left), a channel
      slide without magnetic particles was analyzed at day 6.  Bottom:
      Viability at day 6 as a function of nanoparticle concentration,
      estimated based on intensity gradients as shown in
      Figure~\ref{fig:ParticlesGradient} and the Beer--Lambert law.
      Error bars indicate the standard error for the fraction of dead
      cells and standard deviation for the estimated
      concentration. Cell mortality increases with increasing
      concentration and time up to \SI{40(5)}{\percent} at the highest
      concentration after 6 days of exposure. Cells at a distance of
      \SI{15}{mm} and above, i.e.\ with concentrations below
      \SI{75}{\micro g/mL}, are not noticeably affected.}
    \label{fig:MortalityGradient}
\end{figure}

Figure~\ref{fig:MortalityGradient}, top, shows the viability of HepG2
cells in the presence of bare Fe$_2$O$_3$ nanoparticles with an
average concentration of \qty{100}{\micro g/mL}. If the channel slide
is not positioned on the Halbach array (``control"), the nanoparticles have no effect on  cell viability. This is in
agreement with experimental results for the cylindrical magnet.

If the channel with HepG2 cells is filled when it is positioned on top
of the Halbach array, a gradient forms. If it is kept there for
several days, cell viability strongly decreases. The decrease is
strongest at the start of the channel where the concentration is
highest. From \qty{15}{mm} onwards, there is no difference in the
control.

If we assume that all particles are captured on the array, i.e.\ on the
first half of the microslide, the average concentration doubles to
\qty{200}{\micro g / mL}. Taking the intensities extracted from
Figure~\ref{fig:ParticlesGradient}, we can use the Beer--Lambert model to estimate the concentration as
a function of position. With a single
experiment, we can therefore obtain a first estimate of cell viability
versus concentration. The result is shown in
Figure~\ref{fig:MortalityGradient} (bottom). Cells in areas where the
concentration is below \qty{75}{\micro g/mL} are not noticeably
affected.

%%% Local Variables:
%%% mode: latex
%%% TeX-master: "../paper"
%%% End:

%%% Local Variables:
%%% mode: latex
%%% TeX-master: "../paper"
%%% End:

%%% Local Variables:
%%% mode: latex
%%% TeX-master: "../paper"
%%% End:

% \section{Discussion}
\section{Conclusions}
We constructed magnetic systems that exert a strong force on magnetic
nanoparticles inside a channel slide with a two-dimensional human liver cell
culture. These systems were used to assess the effect of these magnetic
forces on the mortality of mixtures of liver cells and
nanoparticles. The nanoparticles we used were bare Fe$_2$O$_3$ with 
diameters of \num{20} to \qty{40}{nm} and core-shell particles of \qty{100}{nm}
diameter with a magnetite core and a cross-linked dextran shell.

%\cmtr{Part I: magnetic properties}

%\cmtr{Conclusion on cylindrical magnet based on figures~\ref{fig:Field},
 % \ref{fig:FzFxAreaInterest}, \ref{fig:Gradient5Hours},  \ref{fig:ParticlesUniform}}

The first system we investigated exerts a uniform force over the
entire channel via a cylindrical magnet with a diameter of
\qty{70}{mm}. The vector components of the field and force in the
vertical direction are dominant, with a strength in excess of
\qty{300}{mT} and \qty{6}{MN/m^3}, respectively. Over the length of the
30-mm channel, the force strength remains within
\qty{14}{\percent}, and the force direction varies by less than
\qty{1}{\degree}, leading to a concentration gradient that is less than the
measurement uncertainty of \qty{12}{\percent} after \qty{4.3}{h}
exposure. Application of a magnetic field leads to an increased
clustering of particles, both for bare Fe$_2$O$_3$ and core-shell
nanoparticles.

%\cmtr{Conclusion on Halbach magnet based on figures~\ref{fig:FieldGradient},
 % \ref{fig:Halbachgradient}, \ref{fig:trajectory},  \ref{fig:ParticlesGradient}:}

%FieldGradient
The second system is based on a Halbach array that applies non-uniform
fields, the strength of which is on the same order as the field of the large magnet of the
first system. However, the forces are more than one order of magnitude
higher because the individual magnets in the Halbach array are \num{70}
times smaller. Owing to the Halbach configuration, the lateral
forces are both positive and negative and attract particles toward
the transitions. The vertical forces are always downward but vary up
to \qty{30}{\percent}.

% Halbachgradient
When a suspension of bare Fe$_2$O$_3$ is passed over the Halbach
array, particles are captured, and the concentration in the flow
diminishes as the wavefront progresses. As a result, a gradient in
particle concentration is achieved over the length of the Halbach
array.

% trajectory and capture
Calculations show that particles with a radius of less than \qty{20}{nm} and a
remanent magnetization of \qty{1}{T} are trapped above the Halbach
array with an efficiency of less than \qty{7}{\percent}. However, our results show that we are able to capture most of the Fe$_2$O$_3$
particles. Therefore, we conclude that clustering already takes place in
suspension.  Calculations show that clusters with a diameter of \qty{67}{nm} 
and a remanence of \qty{1}{T} are captured with an efficiency of
\qty{51}{\percent}. For cluster radii below this value, capture height
increases linearly with increasing radius. Above this value, 
capture efficiency increases rapidly, and full capture is achieved for
cluster radii greater than \qty{110}{nm}.

% ParticlesGradient
We observed that particles form clusters of  structures with lengths of up to \qty{100}{\um} at the bottom of the channel. At the start of the
Halbach array, these clusters are distributed irregularly. Midway through the
array, however, the distribution becomes more regular. Above the in-plane
magnets of the array, the clusters align along the channel, whereas above the
perpendicular magnets, the clusters become oriented more randomly. Not
all particles are captured by the Halbach array. At the end of the
array, a number of larger clusters can be found. In the remainder of
the channel, chains of particles can be seen aligned along the
channel length.

%\cmtr{Part II: Cell mortality}

Both systems demonstrate cell survival in the presence of
magnetic nanoparticles and strong magnetic field gradients. Application
of a uniform magnetic force reduces the viability of liver cells in
the presence of bare Fe$_2$O$_3$ as well as core-shell nanoparticles
by approximately \qty{50}{\percent}. Cell morphology is not
noticeably dependent on the type of particles, but the presence of the
magnetic field has a strong influence on their distribution. In the magnetic field, the particles agglomerate
into chains. This effect is strongest for the bare Fe$_2$O$_3$
nanoparticles.

%\cmtr{Conclusions on Halbach array based on figure~\ref{fig:MortalityGradient}}

The Halbach array allows us to analyze the cell response as a function
of concentration. In a gradient of Fe$_2$O$_3$ nanoparticles,  cell
mortality depends on the local concentration, ranging from
\qty{40(5)}{\percent} after
six days of exposure at the highest concentration of approximately
\qty{500}{\micro g/mL} to indistinguishable from the control at a concentration
of less than \qty{75}{\micro g/mL}. 

The models presented in this paper offer a solid basis for designing magnetic systems for cell culture studies and particle
trapping. These design methodologies, together with the tools available online,
can be used to modify designs for other channel geometries. The
simulation tools presented here for analyzing particle trajectories  can serve as a fast alternative to finite-element
simulations.\cite{Stevens2021} The capability of creating concentration
gradients may speed up the screening of the effect of magnetic particles
on cell lines. It may also allow other applications, such as cell
migration studies.\cite{Greiner2014}

%%% Local Variables:
%%% mode: latex
%%% TeX-master: "../paper"
%%% End:

\section*{Acknowledgments}
This work was financed by KIST Europe under project number 12008. The
authors would like thank Lilli-Marie Pavka for proofreading the manuscript,
Jonathan O'Connor for design of the microscopy mask and Matthias
Altmeyer and Michiel Stevens for discussions.

\clearpage
\bibliographystyle{apsrev4-2-modified}
%\bibliography{paperbase.bib}

%\cmtr{Additional material: openscad holder, cades input, ImageJ script}
\clearpage
\appendix
\section{Supplementary Material}

\subsection{Optical microscope of stained cells and bare Fe$_2$O$_3$ nanoparticles}
\begin{figure*}
    \centering
    \includegraphics[width=\textwidth]{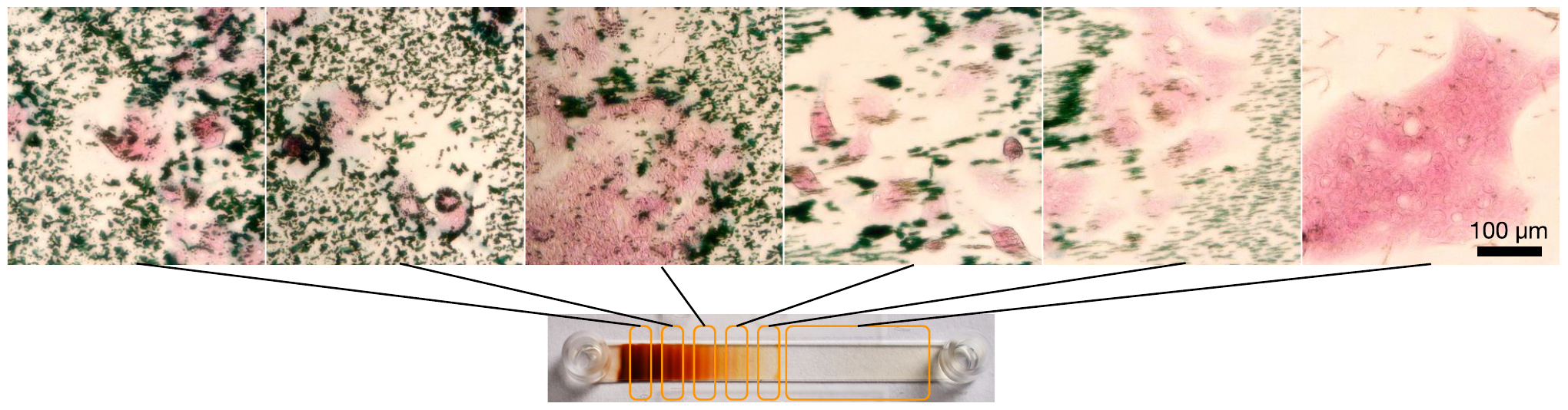}
    \caption{ Optical microscopy of cells (stained red) and bare
      Fe$_2$O$_3$ nanoparticles (stained blue) inside the channel
      slides after 4 days of cultivation. }
    \label{fig:MicroscopyGradient}
\end{figure*}

Figure~\ref{fig:MicroscopyGradient} shows optical microscopy images of
bare Fe$_2$O$_3$ nanoparticles in combination with cells (similar to
Figure~\ref{fig:MicroscopyUniform}) as a function of the position in
the channel. From left to right, the particle concentration decreases, as
indicated by the positions on the image of the slide channel below.
(For clarity, note that this is the image of
Figure~\ref{fig:Halbachgradient} that was at higher concentration and
without cells. The actual particle density is more comparable to
Figure~\ref{fig:ParticlesGradient}.) The cell morphology does not vary
significantly with nanoparticle concentration. The increase in
nanoparticle concentration is clearly visible, but the concentration
is non-uniform over length scales comparable to the cell dimensions.

\subsection{Cades input files}
The compressed folder \texttt{Cades.zip} contains the source files for
the MagMMEMS package. They are used to calculate the magnetic field and forces
of the cylindrical magnet, see Section~\ref{sec:FieldForceCalculations}.

\subsection{Python source files}
The Python source code used to calculate the trajectories of particles
above the Halbach array is available on github:
\href{https://github.com/LeonAbelmann/Trajectory}{https://github.com/LeonAbelmann/Trajectory}

\subsection{3D-print source files}
The OpenScad source files (\texttt{*.scad}) and print files
(\texttt{*.stl}) for the 3D-printed holders are available in the
compressed folder \texttt{3Dprints.zip}.

\subsection{Time-lapse videos}
The time-lapse video \texttt{CoreShell.mp4} shows the change in
concentration over time, see Figure~\ref{fig:Gradient5Hours}. Also shown
is the same experiment for the bare Fe$_2$O$_3$ suspension,
\texttt{Fe2O3.mp4}, where the formation of gas bubbles can be
observed.

\subsection{Gradient filling}
The formation of the concentration gradient can be observed in the
video \texttt{GradientFilling.mp4}. The end result is shown in Figure~\ref{fig:Halbachgradient}.

\subsection{High resolution stiched images}
Figures \texttt{FullGradientFe2O3.png} and
\texttt{FullGradientCoreShell.png} show high-resolution versions of the
images in Figure~\ref{fig:ParticlesGradient}.

%%% Local Variables:
%%% mode: latex
%%% TeX-master: "../paper"
%%% End:

\end{document}